\documentclass[a4paper,10pt,reqno]{amsart}
\usepackage[utf8]{inputenc}
\usepackage[%
    style=numeric-comp, hyperref,backend=biber, doi=false,url=false,
    sorting=none,backref=true, maxbibnames=7, firstinits=true]{biblatex} 
\renewbibmacro{in:}{}
\usepackage[foot]{amsaddr}
\usepackage{amsmath}
\usepackage{amssymb}
\usepackage{bbm}
\usepackage{bm}
\usepackage{xcolor}
\numberwithin{equation}{section}
\numberwithin{figure}{section}

\usepackage{graphicx}
\usepackage{epstopdf}
\usepackage{hyperref}
\hypersetup{colorlinks=true,linktoc=page,linkcolor=blue,citecolor=red,urlcolor=cyan}

\bibliography{bibliografia}

\usepackage[english]{babel}

\newcommand{\hu}{\hspace{1cm}}

\newcommand{\paras}[1][x]{\bm{#1}^\parallel}
\newcommand{\para}[1][x]{{{#1}^{\parallel}}}
\newcommand{\perpe}[1][x]{{#1}^D}
\newcommand{\dime}{D}
\newcommand{\dk}[1][k]{\frac{{\rm d} #1}{(2\pi)}}
\newcommand{\dx}[1][x]{{{\rm d} #1}}

\newcommand{\mathi}{{\text{i}}}

\newcommand{\ampli}{\zeta}
\newcommand{\ff}{F}
\newcommand{\Kdelta}[1][]{K_{\rm delta}}
\newcommand{\Kzeta}[1][\ampli]{K_{#1}}
\newcommand{\Keta}[1][\zeta]{K_{#1}}

\newcommand{\opeinho}{A_{}}
\newcommand{\pos}[1][]{L_{#1}}

\newcommand{\tildex}{\tilde x}
\newcommand{\tildey}{\tilde x'}

\usepackage{xparse}

\NewDocumentCommand{\dkpara}{O{k} O{\dime}}{\frac{{\rm d} {#1}^{\parallel}}{(2\pi)^{#2}}}
\NewDocumentCommand{\dkd}{O{k} O{\dime}}{ \frac{{\rm d}^{#2} {#1}}{(2\pi)^{#2}}}
\NewDocumentCommand{\dxd}{O{x} O{\dime}}{ {\rm d}^{#2} {#1} }
\NewDocumentCommand{\ad}{O{} O{}}{ \bigl \langle {#1}\bigr\rangle_{{\rm ad}{#2}} }

\NewDocumentCommand{\pathi}{O{} O{} O{} O{}  O{} O{x}}{ \int_{#6(#3)=#2}^{#6(#5)=#4} \mathcal{D}{#6} {#1}}
\NewDocumentCommand{\pathidx}{O{} O{} O{} O{} O{}  O{} O{x}}{ \int_{#6(#3)=#2}^{#6(#5)=#4} \mathcal{D}{#7} {#1}}
\NewDocumentCommand{\pathifree}{O{} O{} O{} O{} O{} O{x}}{\pathi[\mathcal{D}p\, e^{-\int_{#3}^{#5} \dx[t] \big(p^2-\mathi p \dot{x} \big)}][#2][#3][#4][#5][#6]}
\makeatletter
\def\@tocline#1#2#3#4#5#6#7{\relax
  \ifnum #1>\c@tocdepth % then omit
  \else
    \par \addpenalty\@secpenalty\addvspace{#2}%
    \begingroup \hyphenpenalty\@M
    \@ifempty{#4}{%
      \@tempdima\csname r@tocindent\number#1\endcsname\relax
    }{%
      \@tempdima#4\relax
    }%
    \parindent\z@ \leftskip#3\relax \advance\leftskip\@tempdima\relax
    \rightskip\@pnumwidth plus4em \parfillskip-\@pnumwidth
    #5\leavevmode\hskip-\@tempdima
      \ifcase #1
       \or\or \hskip 2em \or \hskip 4em \else \hskip 6em \fi%
      #6\nobreak\relax
    \dotfill\hbox to\@pnumwidth{\@tocpagenum{#7}}\par
    \nobreak
    \endgroup
  \fi}
\makeatother
\title[Local Neumann semitransparent layers]{Local Neumann semitransparent layers: resummation, pair production and  duality }

\author{N. Ahmadiniaz$^{1}$}
\address{$^1$ Helmholtz-Zentrum Dresden-Rossendorf, Bautzner Landstraße 400, 01328 Dresden, Germany.}

\author{S.~A.~Franchino-Vi\~nas$^{1,2,3}$}

\address{$^2$ Departamento de F\'isica, Facultad de Ciencias Exactas,
Universidad Nacional de La Plata, C.C.\ 67 (1900), La Plata, Argentina.}

\email{$^3$ \href{mailto:s.franchino-vinas@hzdr.de}{s.franchino-vinas@hzdr.de} }

\author{L. Manzo$^{2,4}$}

\address{$^4$ Instituto de F\'isica La Plata, CONICET and Universidad Nacional de La Plata, C.C.\ 67 (1900), La Plata, Argentina.}

\author{F.~D.~Mazzitelli$^{5,6}$}

\address{$^5$ Centro At\'omico Bariloche,  CONICET,
Comisi\'on Nacional de Energ\'\i a At\'omica, R8402AGP Bariloche, Argentina.}

\address{$^6$
Instituto Balseiro, Universidad Nacional de Cuyo, R8402AGP Bariloche, Argentina. }

\begin{document}

\maketitle

\begin{abstract}
We consider local semitransparent Neumann boundary conditions for a quantum scalar field as imposed by  a quadratic coupling to a source localized on a flat codimension-one surface. 
Upon a proper regularization to give  meaning to the interaction, we interpret the effective action as a theory in a first-quantized phase space.
We compute the relevant heat-kernel to all order in a homogeneous background and quadratic order in perturbations, giving a closed expression for the corresponding effective action in $D=4$. In the dynamical case, we analyze the pair production caused by a harmonic perturbation and a Sauter pulse.
Notably, we prove the existence of a strong/weak duality that links this Neumann field theory to the analogue Dirichlet one.

\end{abstract}

\newpage

\tableofcontents

\newpage

\section{Introduction}

One of the major successes of Quantum Field Theory (QFT) has been the prediction of the Casimir effect~\cite{Casimir:1948dh},
which builds a bridge between the world of macroscopic media and that of quantum effects~\cite{Bordag:2009zz, Lamoreaux:2005gf, Milton:2001yy, Kirsten:2001wz, Mostepanenko:1990ceg}.
In a first approximation, the bodies may be modelled as perfect conductors and thus implemented as boundary conditions on the electromagnetic field~\cite{Casimir:1948dh}.
Of course, this is not enough to describe the always improving experimental results 
\cite{Klimchitskaya:2021qxk, Bimonte:2021sib}.
Moreover, this simplification is thought to be the root of some theoretical issues that include the divergence of the energy density at the boundaries~\cite{Deutsch:1978sc, Graham:2003ib}
and a possible ill-definition of the vacuum self-energy~\cite{Estrada:2012yn}.

A way to obtain more realistic models, is to incorporate information  of the bodies' bulk, as in Lifshitz theories~\cite{Lifshitz:1956zz}. 
In recent times, this has been done by modelling electromagnetic properties as external (classical) smooth fields 
\cite{Graham:2002xq,Efrat:2021kjr,Franchino-Vinas:2021lbl,Shayit:2021kgn,Parashar:2018pds, Fulling:2018qcn, Milton:2016sev, Leonhardt:2011zz,Griniasty:2017iix,Griniasty:2017ofc,Murray:2015tim}.

Another possibility regards the substitution of perfect Dirichlet or Neumann boundary conditions by more general ones.
In the Dirichlet case, one can mention for example semitransparent boundary conditions~\cite{Bordag:1992cm,Cavero-Pelaez:2006niz, Cavero-Pelaez:2006wkx,Franchino-Vinas:2022}. 
These can be seen as a special case of the ones described in the previous paragraph,
where potentials get localized in a thin shell and allow the recovery of perfect boundary conditions as a special limit of the coupling.

Following these lines one can also consider local semitransparent conditions,
in which the coupling (or background field living on the thin shell) becomes spacetime dependent~\cite{Bordag:2004rx, Fosco:2019lmw}.
This allows the study of interesting phenomena and cases, such as particle creation~\cite{Franchino-Vinas:2020okl} and smooth nonplanar geometries (say gratings~\cite{Oue:2021PRA}) in the effective-medium approach~\cite{Kidway:2012}.

A by far less studied case regards the analog for Neumann boundary conditions. 
It has been shown in Ref.~\cite{Fosco:2009ic} that Neumann semitransparent (or imperfect) boundary conditions 
for a scalar field
can be modelled by the operator of quantum fluctuations
\begin{align}\label{eq:operator}
\Delta_{\ampli}:&=-\partial^2_x-\ampli \partial_x \delta(x) \partial_x ,
\end{align}
where $\ampli\in\mathbb{R}$ is the coupling constant. From a mathematical point of view,
the problem at hand mixes several interesting ingredients. 
Indeed, it is contained in the so-called  four-parameter family of point interactions~\cite{Albeverio:1988}, 
i.e. it arises in the study of self-adjoint extensions 
of the one-dimensional second-derivative operator acting on  functions appropriately defined on $\mathbb{R}/\{0\}$. 
In this context the interaction was baptized  $\delta'(x)$, leading to some misunderstandings in the literature.
This point was already explained in the paper by Šeba \cite{Seba1986}, 
which showed that the correct interpretation of this self-adjoint extension is in terms of a $\partial \delta \partial$ potential with a renormalized potential, 
as also discussed in \cite{Albeverio2013ARS, Fosco:2009ic, Grosse:2004rp}.
In particular,  $\Delta_{\ampli}$ is not related to the proposal in \cite{Gadella:2011}, where a true generalized potential is considered.

One formal way to introduce the operator $\Delta_{\ampli}$ is through the implementation of boundary conditions \cite{Albeverio:1988, Asorey:2004kk}.
In this case, the action of $\Delta_{\ampli}$ should be understood as implying the boundary conditions
\begin{align}
 \phi(0^+)-\phi(0^-)=\ampli \phi'(0),
\end{align}
where $0^{\pm}$ denotes the left/right limits to zero, and the derivative should be regularized for example as $\phi'(0)=\frac{\phi'(0^+)+\phi'(0^-)}{2}$,
see \cite{Grosse:2004rp,Albeverio:1988}. This interpretation through boundary conditions has been pursued recently in \cite{Cavero-Pelaez:2020pxy, Gadella:2019ujd},
where $\delta$-$\delta'$ structures have been considered.

In the present paper we will focus on a scalar field living in a $\dime$-dimensional flat space.
It will satisfy local Neumann semitransparent boundary conditions on a flat thin shell, 
i.e. we will upgrade $\ampli$ to a spacetime-dependent coupling (or field living on the shell). 
In particular, we will follow the idea that such a field may develop small fluctuations $\eta$ around a mean (or vacuum expectation) value $\ampli$,
so that an exact treatment in $\ampli$ but a perturbative one in $\eta$ should provide rich physical information. 
We thus begin our exposition in Sec.~\ref{sec:model} with a description of the relevant QFT, consisting of a scalar field interacting with a classical background.
In Sec.~\ref{sec:worldline} we explain how to compute the heat-kernel of $\Delta_{\ampli}$ 
to all order in the coupling $\ampli$ employing the Worldline Formalism, i.e. a path-integral approach. 
In doing so, we devise a tailor-made regularization which allows to perform 
 in Sec.~\ref{sec:inhomogeneous} a perturbative computation for spacetime-dependent contributions, which is also exact in the constant background.

 Then, in Sec.~\ref{sec:ea} we analyze the effective action in a four-dimensional setting,
showing that the renormalized theory is dual to the problem of local (Dirichlet) semitransparent boundary conditions.
Afterwards, we study in Sec.~\ref{sec:dynamical} the effect of pair production in our system by considering time-dependent background fields,
an analogue of the widely studied dynamical Casimir effect. 
In particular, as examples we consider  a harmonic perturbation and a Sauter pulse.
Finally, in Sec.~\ref{sec:conclusion} we state our conclusions.

We leave necessary technical information to the appendices, where we compute the Worldline generating function relevant to our current problem (App.~\ref{app:vev}), 
compute integrals involving chained free heat-kernels (App.~\ref{app:integrals};  others containing also  Hermite polynomials are included in App.~\ref{app:integral_hermite}) and calculate in closed form a series of Hermite polynomials (App.~\ref{app:series_hermite}).
We use Planck units so that $\hbar$ and $c$ are taken to be unity. 
It will prove convenient to split $\dime$-dimensional coordinates $x$ into the coordinate perpendicular to the shell, $\perpe[x]$, and those $\dime-1$ parallel to it, $\para[x]$; in Minkowski space, the set of spacelike coordinates of the latter is denoted by $\paras[x]$. We define $\tilde x:= x-\pos$.

%\colred{We may also cite some similar problems: point delta in more dimensions, spinors, etc.}
\iffalse 

\ref{app:vev} \ref{app:rescaling}

\fi

\section{The model}\label{sec:model}
We begin by considering a scalar quantum field $\varphi$ living on a $\dime$-dimensional Euclidean flat spacetime;
it interacts with an external (classical) scalar field $\eta$ that lives on a plate 
according to the following action:
\begin{align}\label{eq:action}
 S:=\frac{1}{2}\int \dxd[x][\dime] \, \Big[ (\partial \varphi)^2 +m^2\varphi^2 - \big(\eta(\para[x])+\ampli\big) \delta(\perpe-\pos ) (\perpe[\partial]\varphi)^2\Big].
\end{align}
In this action $m$ is the mass of the  field $\varphi$ and
we have splitted the coordinates into the direction perpendicular to the plate ($\perpe[x]$) and those parallel to it ($\para[x]$).
Additionally, the plate is placed at $\perpe=\pos$ and $\ampli$
may be understood as a mean value of the field $\eta$ over the plate. From a physical point of view, the classical field describes the properties of the thin plate that are relevant in the interaction with the quantum field.
Notice also that $\eta$ has dimensions of length, independently of the dimension $\dime $ of the spacetime.

The interpretation of this action in Eq.~\eqref{eq:action} is more evident once we consider a homogeneous configuration for which $\eta(\para[x]) \equiv 0$ \cite{Fosco:2009ic}:
in the limit $\ampli\to \infty$ one expects to obtain two semi-spaces, on whose boundaries $\varphi$ satisfies Neumann boundary conditions, 
very much akin to the generation of Dirichlet boundary conditions through delta potentials\footnote{We will see in Sec. \ref{sec:worldline_hk} that the correct limit is indeed $\ampli\to \infty$ and not the naive one $\ampli\to -\infty$.}.
In this sense, for finite $\ampli$ the action \eqref{eq:action} can be interpreted as imposing Neumann semitransparent boundary conditions on the field $\varphi$;
allowing for inhomogeneous field configurations $\eta(\para[x])$, we will refer to local Neumann semitransparent boundary conditions.

Following the usual way, upon a path-integral quantization one can obtain the generating functional $\mathcal{Z}$ for this system,
\begin{align}
 \mathcal{Z}[J]:=\int \mathcal{D}\varphi\, e^{-S+J\varphi}.
\end{align}
Afterwards, 
%employing a Legendre transform to the generating functional of connected correlation functions, $W:=-\log \mathcal{Z}$, 
integrating out the scalar field $\varphi$, we can write the effective action $\Gamma$ in terms of the operator of quantum fluctuations $A$;
a direct computation gives
\begin{align}
 \Gamma_{\rm one-loop}&= \frac{1}{2} \operatorname{Log} \operatorname{Det} \left( A\right),
 \\
 A(-\mathi \partial,x,\eta):&=-\partial^2+m^2+ \big( \eta(\para[x])+\ampli \big)\perpe[\partial] \left[ \delta(\perpe-\pos) \perpe[\partial]\right],
 \label{eq:quantum_fluctuations}
\end{align}
where $\Gamma_{\rm one-loop}$ denotes the quantum contributions to $\Gamma$.
We will assume that $\eta+\ampli\geq0$, such that $A$ admits only a continuum spectrum.
A peculiarity is that the  operator $A$  may be interpreted as a Schrödinger operator with a derivative-dependent potential, 
different from the more diffused case of an only position-dependent potential\footnote{Another way to look at the 
problem is to focus on the similarities that the operator $A$ has with Laplace-Beltrami operator with a singular metric. We are not going to pursue this way in this article.}. 
As we will see in the next section, this admits an interpretation of boundary conditions in phase space.

On physical grounds, we expect two interesting regimes for our system: one in which $\eta$ develops small fluctuations around a constant background, admitting thus an expansion in powers of the fluctuations, and one in which topology plays an important role, such that $\eta$ should be considered to all orders. 
In the following we will consider the former, leaving the latter for future studies.

\section{A path integral approach: the Worldline Formalism in phase space}\label{sec:worldline}
One simple way to compute the quantum contributions to the effective action is to employ the well-known equivalence
between $\text{Log}\,\text{Det}$ and $\text{Tr}\,\text{Log}$, as well as Schwinger's propertime trick (or Frullani's 
representation for the logarithm of a quotient~\cite{Jeffreys}).
In this way we may recast 
\begin{align}\label{eq:EA_trlog}
\begin{split}
\Gamma_{\rm one-loop} &= \frac{1}{2}\operatorname{Tr} \operatorname{Log} A
 = -\frac{1}{2}\int_0^{\infty} \frac{\dx[T]}{T} \operatorname{Tr} K_A(x,x';T) ,
\end{split}
\end{align}
where the heat-kernel $K_A(x,x';T):=\langle x \vert e^{-T A} \vert x'\rangle$ has appeared in a natural way. 

From the formal side, the study of the spectral functions of operators with singular operators or generalized boundary conditions has attracted much attention in recent years~\cite{Falomir:2020tjw, Kirsten:2005bh, Falomir:2003vw, Gilkey:2002nv, Branson:1999jz,Bordag:1999ed}.
One efficient way to perform this kind of computations is through the Worldline Formalism, in which  one interprets the heat kernel in terms of path integrals in a first quantization procedure. 
Indeed, one may notice that the arguments of $A$ are momentum and position operators in a first quantization, 
realized as $( \hat p,\hat x) \to (-\mathi \partial, x)$, so that $e^{-T A}$ can be understood as the evolution operator in imaginary time $t=\mathi T$. 

The Worldline Formalism has been successfully applied to several problems, see the book \cite{Bastianelli:2006rx}, the reviews \cite{Schubert:2001he,ahmadiniaz-review} and references therein. 
In particular, it has recently proved useful in three situations that  are relevant to the present computation:
in phase space, where it has been applied to investigate noncommutative quantum field theories \cite{Franchino-Vinas:2021bcl, Franchino-Vinas:2018jcs, Bonezzi:2012vr} and Berry phases \cite{Copinger:2022jgg},  in the analysis of singular potentials and metrics \cite{Franchino-Vinas:2020okl, Corradini:2019nbb, Vinas:2010ix} and in the  study of
boundaries \cite{ Bastianelli:2006hq, Bastianelli:2007jr, Edwards:2021cyp} (see also \cite{Clark:1980xt, Farhi:1989jz, Carreau:1991yx} for related path integral approaches).

As a first step, we will perform an all-order computation with a constant background, setting $\eta\equiv 0$. 
It should be clear that under such assumption one can disentangle the contribution in the $\dime$th direction,
yielding the remaining components just a free path integral. 
Therefore, for the rest of this section we may simply work in a one-dimensional setup.
Additionally, we will consider the massless case, 
given that the mass may be directly included in the heat-kernel at the end of the computations, simply by adding a factor $e^{-m^2T}$. 

Taking these considerations into account, we follow the Worldline Formalism approach to compute the transition amplitudes, 
so that the relevant heat-kernel may be  written as
\begin{align}\label{eq:master_hk}
 \Kzeta(x,x';T) :&= \pathi[\mathcal{D}p \, e^{-\int_0^T\dx[t] \,\left[ p^2(t) -\mathi p(t)\dot{x}(t) -\ampli \left( \delta(x-\pos) p^2 +\frac{1}{4}\delta''(x-\pos)\right)\right] }][x][0][x'][T] .
\end{align}
In obtaining the master equation \eqref{eq:master_hk} one needs to Weyl-order the operator $A$, 
which involves employing the commutation relation between derivatives and coordinates to render the operator symmetric in terms of $p$ and $x$, see \cite{Berezin:1971jf, Bastianelli:2006rx}; 
this is the origin of the $\delta''$ contribution and one of the reasons why previous attempts to compute the heat-kernel of the operator $A$ may have failed \cite{Grosche:1994uv}. 
As a matter of fact, one should keep in mind that this additional $\delta''$ term does not enter in any way in the original operator; instead, it just plays a role in the path integral interpretation.

One further important point is that the path integral over the momentum variables has no boundary conditions.
Since from the QFT point of view we are interested in the trace of the heat-kernel, we could have
imposed periodic boundary conditions on the phase space path integral. However, such a choice would not allow the study of the heat-kernel out of the diagonal that we are going to undertake in the following section. 

\subsection{The heat-kernel \texorpdfstring{$\Kzeta$}{}}\label{sec:worldline_hk}
We will compute the heat-kernel in a perturbative fashion, expanding the result \eqref{eq:master_hk} in the coupling parameter $\ampli$. 
In this way we obtain
\begin{align}
 \begin{split}\Kzeta(x,x';T) &=  \pathi[\mathcal{D}p \, ][x][0][x'][T] e^{-\int_0^T\dx[t] \,\left[ p^2(t) -\mathi p(t)\dot{x}(t) \right] } 
 \\
 &\hspace{-1.cm} \times\sum_{n=0}^{\infty} \frac{\ampli^n}{n!} \,
 \prod_{j=1}^n \Bigg\{\int_0^T  \dx[t_j] \left[ \delta(x(t_j)-\pos) p^2(t_j)+\frac{1}{4}\delta''(x(t_j)-\pos) \right]\Bigg\}
 .
 \end{split}
\end{align}
To proceed further we  compute the momentum integrals, which are Gaussian upon introducing a source denoted by $j$:
\begin{align}
\begin{split}
\int \mathcal{D}p\, e^{-\int_0^T \dx[t]\left(p^2 -\mathi \dot x p\right)} p^2(t_1)
&=\mathcal{N} \frac{\delta^2}{\delta {j}(t_1)^2} e^{\frac{1}{4} \int_0^T \dx[t] (j+\mathi \dot x)^2  } \Big\vert_{j=0}.
\end{split}
\end{align}
The factor $\mathcal{N}$ is a normalization constant whose only role is to be fixed when determining the value of the free path integral and therefore may be safely dismissed.
Having recasted every momentum as a variation in $j$, one can simplify the problem noting that 
$\delta''(x-\pos)=\partial^2_{\pos} \delta(x-\pos)$; this enables us to use the Dirac deltas to impose constraints on the paths
 as following:
\begin{align}
 \begin{split}
&\Kzeta(x,x';T) = \sum_{n=0}^{\infty} \left(\frac{\ampli }{4}\right)^n \int_S \prod_{l=1}^n \left[ \dx[t_l]\left( 4 \left(\frac{\delta}{\delta {j_l}}\right)^2+ \partial_{\pos[l]}^2 \right)\right]
\\
&\hu\times\pathi[\,e^{\int_{0}^{t_1} \dx[t]\, \frac{(j+\mathi \dot x)^2}{4}} ][x][0][\pos[1]][t_1] \cdots \pathi[\,e^{\int_{t_n}^T \dx[t]\, \frac{(j+\mathi \dot x)^2}{4} }   ][\pos[n]][t_n][x'][T] \Big\vert_{j=0} ,
 \end{split}
\end{align}
where for any $t$-dependent quantity $X(t)$ adding a subindex means $X_l:=X(t_l)$ and
the integral over the intermediate times has been ordered, such that
\begin{align}
 \int_S \prod_{j=1}^n  \dx[t_j] :&= \int_0^T \dx[t_n] \int_0^{t_n} \dx[t_{n-1}] \cdots \int_0^{t_2} \dx[t_1] .
\end{align}
Additionally, we have introduced one position variable ($L_i$) for each insertion of the potential, in order to avoid undesired mixing of the derivatives; 
as we will see shortly, this will bring its own benefits.

In this way the computation is reduced to a chain of partition functions (or generating functionals), which can be readily computed as in App. \ref{app:vev}. 
Using those results we get
\begin{align}
   \begin{split}
\Kzeta(x,x';T) 
&= \sum_{n=0}^{\infty} \left(\frac{\ampli }{4}\right)^n     \int_S \prod_{l=1}^n \left[ \dx[t_l]\left( 4 \left(\frac{\delta}{\delta {j_l}}\right)^2+ \partial_{\pos[l]}^2 \right)\right]
\\
&\hu\hu\hu\times\prod_{m=0}^{n} \left.\left[ \frac{e^{\frac{1}{4\Delta t_m} \left (\mathi \Delta x_m+ \int_{t_{m}}^{t_{m+1}} \dx[t]\, j(t) \right)^2}}{\sqrt{4\pi\Delta t_m}}   \right]\right\vert _{j=0},
 \end{split}
\end{align}
where we have introduced the natural notation for the intermediate displacements $\Delta x_{m}:=x_{m+1}-x_{m}$ 
and intermediate periods $\Delta t_m:= t_{m+1}-t_m$;
correspondingly we define $x_0:=x$, $x_{n+1}:=x'$, $t_0:=0$, $t_{n+1}:=T$ and  for $i=1,\cdots n$ we set $x_i:=\pos[i]$.
Even if at first sight the computation of the variations and derivatives may seem a hard task,
the implementation of a Hubbard--Stratonovich transformation to linearize the problem allows a direct computation:
\begin{align}
   \begin{split}
&\Kzeta(x,x';T) 
\\
&= \sum_{n=0}^{\infty} \left(\frac{\ampli }{4}\right)^n     \int_S \prod_{l=1}^n \left[ \dx[t_l]\left( 4 \left(\frac{\delta}{\delta {j_l}}\right)^2+ \partial_{\pos[l]}^2 \right)\right]
\\
&\hu\times \int_{-\infty}^{\infty}\prod_{m=0}^{n} \left. \left[\dk[k_m] e^{- \Delta t_m k_m^2 +k_m \left (\mathi \Delta x_m+ \int_{t_{m}}^{t_{m+1}} \dx[t]\, j(t) \right)}   \right] \right\vert_{j=0}
\\
&= \frac{(-1)}{2 \sqrt{\pi}}\sum_{n=0}^{\infty} \left(\frac{\ampli }{8\sqrt{\pi}}\right)^n     
\int_S \left[ \prod_{l=1}^{n} \dx[t_l] \right]   \frac{ \Delta x_0 \Delta x_n}{\Delta t_0^{3/2} \Delta t_n^{3/2}} \left[ \prod_{m=1}^{n-1} \frac{(2\Delta t_m-\Delta x_m^2)}{\Delta t_m^{5/2}} \right] \prod_{p=0}^{n} e^{-\frac{\Delta x_p^2}{4\Delta t_p}}.
 \end{split}
\end{align}

Notice that had we set $L_i\equiv L$ at this point, the integrals in the intermediate times would have become divergent. 
Indeed, it has been shown in the past that the problem at hand requires a renormalization of the coupling~\cite{Seba1986, Fosco:2009ic} 
or, alternatively, an extension of the  the potential  to let it act on more general functions~\cite{Albeverio:2000}. 
This feature is shared by some related problems, such as point  interactions in three dimensions~\cite{Berezin:1961, Albeverio:1988}.
In our case,  the regularization is already implemented by the separation of the intermediate points, 
which in physical terms corresponds to the idea of a plate with finite width suggested in~\cite{Fosco:2009ic}. 
The limit $\pos[m]\equiv \pos$ will thus be left to the last stage of the computation.

As explained in App. \ref{app:integrals},  
one can use the results in \cite{Franchino-Vinas:2020okl} to perform the integrals in the intermediate times one by one.
Considering the first cases, one realizes that the result for arbitrary $n$ involves Hermite polynomials;
using the general result of App. \ref{app:integral_hermite}, we obtain 
\begin{align}\label{eq:hk_perturbative_coefficients}
 \Kzeta(x,x'; T) =:\sum_{n=0}^\infty  \ampli^n \Kzeta^{(n)}(x,x'; T),
\end{align}
where the $n$th order coefficient in this expansion reads
\begin{align}\label{eq:hk_zeta_n}
 \Kzeta^{(n)}(x,x';T)&=
 -\frac{(-1)^n T^{-\frac{n}{2}-\frac{1}{2}} \text{sign}(\tildex) \text{sign}(\tildey) e^{-\frac{(\left|\tildex\right| +\left|\tildey \right| )^2}{4 T }} H_n\left(\frac{\left| \tildex\right| +\left| \tildey \right| }{2 \sqrt{T }}\right)}{2 \sqrt{\pi }}.
\end{align}
In this expression we have employed the sign function $\text{sign}(\cdot)$ and defined the displaced variables,
\begin{align}\label{eq:xtilde}
\tilde x:=x-\pos.
\end{align}
We can perform a resummation in Eq.~\eqref{eq:hk_perturbative_coefficients} by using the method described in App.~\ref{app:series_hermite};
in this way we get
a closed result in terms of the complementary error function $\text{erfc}(\cdot)$:
\begin{align}\label{eq:HK_integral_analytic} 
 \begin{split}
  \Kzeta(x,x';T)&= \frac{1 }{\sqrt{4 \pi T }}  e^{-\frac{(\tildex-\tildey)^2}{4 T }}+ \frac{1 }{ \sqrt{4 \pi T }} \text{sign}(\tildex) \text{sign}(\tildey) e^{-\frac{(|\tildex|+|\tildey|)^2}{4 T }}
  \\
  &\hu-   \frac{1}{ \ampli } \text{sign}(\tildex) \text{sign}(\tildey)e^{\frac{2 \ampli  (\left| \tildex\right| +\left| \tildey\right| )+4T }{ \ampli ^2}}\text{erfc}\left(\frac{\ampli  (\left| \tildex\right| +\left| \tildey\right| )+4T }{2 \sqrt{T } \ampli }\right).
 \end{split}
\end{align}

The expression \eqref{eq:HK_integral_analytic} coincides with the one obtained in \cite{Grosche:1994uv} 
from a fermionic path integral.
One further confirmation of the correctness of our result can be obtained from its trace.
Integrating the heat-kernel \eqref{eq:HK_integral_analytic} over the whole space we get
\begin{align}\label{eq:hk_trace}
 \int \dx[x]\,\Kzeta[\ampli](x,x;T)=\frac{V_{}}{(4\pi T)^{1/2}}+\frac{1}{2}e^{\frac{4 T}{\zeta ^2}} \text{erfc}\left(\frac{2 \sqrt{T}}{\zeta }\right),
\end{align}
where $V_{}$ denotes the volume (length) of the whole space.
Using this formula as point of departure, we can analyze the small and large coupling regimes by using the following expansions:
\begin{align}
 e^{\frac{4 T}{\zeta ^2}} \text{erfc}\left(\frac{2 \sqrt{T}}{\zeta }\right)
 &=\begin{cases}
    \frac{1}{2}-\frac{2 \sqrt{T}}{\sqrt{\pi } \zeta }+\frac{2 T}{\zeta ^2}-\frac{16 T^{3/2}}{3 \sqrt{\pi } \zeta ^3}+\mathcal{O}\left({\zeta }^{-4}\right),
    \\
    \frac{\zeta }{4 \sqrt{\pi } \sqrt{T}}-\frac{\zeta ^3}{32 \left(\sqrt{\pi } T^{3/2}\right)}+\mathcal{O}\left(\zeta ^4\right).
   \end{cases}
\end{align}
In particular, it is immediate to see that we recover the free ($\ampli=0$) and the Neumann  ($\ampli\to\infty$) cases~\cite{Vassilevich:2003xt}.

As a final comment, recall that we are considering a positive $\ampli$. The case with $\ampli<0$ is subtler, given that a bound state with energy $E_{\rm b}=-\frac{4}{\ampli^2}$ arises~\cite{Grosche:1994uv}. 
In order to obtain the heat-kernel trace for negative coupling one should then not only change the sign of $\ampli$ in Eq. \eqref{eq:hk_trace} but also add a further contribution
coming from the bound state, which reads
\begin{align}
 \int \dx\, \Delta\Kzeta[\ampli](x,x;T) = e^{-T E_{\rm b}} =e^{\frac{4T}{\ampli^2}}.
\end{align}
Interestingly, the appearance of the bound state is already signaled as a nonperturbative factor in Eq.~\eqref{eq:hk_trace}, which for $\ampli<0$ generates a strong divergence as $\ampli\to0$. 
This  reminds us of similar effects in resurgence theory, where information about the nonperturbative sector is stored in the perturbative results, see \cite{Fujimori:2022lng,Dunne:2021ctj,Kamata:2021jrs, Dunne:2012ae} and references therein.
A more detailed analysis of this fact will be left to a future publication.

\subsection{Relation with the heat-kernel of a Dirac delta potential}\label{sec:hk_duality}
One fundamental remark concerns the relation of this heat-kernel with a similar problem, viz. that involving a Dirac delta potential. 
If one defines the operator 
\begin{align}
 A^{\rm delta}:&=-\frac{{\rm d}^2}{{\rm d} x^2}+ \lambda \delta(x-\pos) ,
\end{align}
then its heat-kernel has been shown to be \cite{Bauch_1985, Gaveau_1986, Franchino-Vinas:2020okl}
\begin{align}
 \begin{split}\label{eq:hk_delta}
&\Kdelta(x,x';T;\lambda)
 \\
 &=\frac{e^{-\frac{(x-x')^2}{4 T}}}{\sqrt{4\pi T}}-\frac{\lambda  }{4} e^{\frac{1}{4} \lambda  \big[\lambda  T+2 (\left| x-L\right| +\left| x'-L\right| )\big]} \text{erfc}\left(\frac{\left| x-L\right| +\left| x'-L\right| +\lambda  T}{2 \sqrt{T}}\right).
 \end{split}
\end{align}
This means that we may recast our result \eqref{eq:HK_integral_analytic}  as
\begin{align}
 \begin{split}\label{eq:HK_duality}
\Kzeta(x,y;T)&=\frac{\left(1+ \text{sign}(\tildex) \text{sign}(\tildey) \right) }{2 \sqrt{4 \pi T }}  \left( e^{-\frac{(x-x')^2}{4T} }+ e^{-\frac{(\tildex+\tildey)^2}{4 T}}\right)
 \\
 &\hu+{\text{sign}(\tildex) \text{sign}(\tildey)} \left[ \Kdelta(x,x';T;4\ampli^{-1})-\frac{e^{-\frac{(x-x')^2}{4 T}}}{\sqrt{4\pi T}} \right].
 \end{split}
\end{align}
This relation is similar in nature to the Fermi--Bose duality introduced by Girardeau~\cite{Girardeau:1960} and then further developed by Cheon and Shigehara~\cite{Cheon:1998iy} for a gas of interacting particles. 
However, our case differs from theirs, inasmuch as one can verify that $\Kzeta$ possesses no defined symmetry under $\tilde x \to -\tilde x$
and consequently no statistics can be clearly assigned.  
Alternatively, 
projecting our heat-kernels $\Kzeta$ and $\Kdelta$ respectively to the space of antisymmetric and symmetric functions around $\pos$, one can obtain the Fermi--Bose duality at the level of heat-kernels.

Importantly, the similarity between the heat-kernels involved  in~\eqref{eq:HK_duality} gets enhanced once we consider their diagonal, since then some  partial cancellations occur and the sign functions simplify.
As we will see in Secs.~\eqref{sec:map} and \eqref{sec:QFT_duality}, the relation will be upgraded to a map between heat-kernels  for the case of inhomogeneous backgrounds and will entail a duality at the level of QFTs, linking  Neumann and Dirichlet  local semitransparent boundary conditions.

%%%%%%%%%%%%%%
%%%%%%%%%%%%%%
%%%%%%%%%%%%%%
%%%%%%%%%%%%%%
%%%%%%%%%%%%%%
\section{An expansion for local Neumann semitransparent boundary conditions }\label{sec:inhomogeneous}
Let us now turn our attention to the operator in Eq. \eqref{eq:quantum_fluctuations}. 
As precedently commented, we will split the classical background into a constant $\ampli$ plus a small perturbation $\eta(\para[x])$ that may depend on the $(\dime-1)$ coordinates parallel to the plate.
The idea is to perform an expansion in $\eta(\para[x])$,   keeping the full dependence on $\ampli$.
Following the lines in the preceding section one can obtain a Worldline formula for the corresponding heat-kernel:
\begin{align}\label{eq:inhomo_HK}
 \Keta(x,y;T;\eta] &= \int_{x(0)=x}^{x(T)=y}\hspace{-0.5cm}\mathcal{D}x\mathcal{D}p \, e^{-\int_0^T\dx[t] \,\left[ p^2(t) -\mathi p(t)\dot{x}(t) - (\ampli+\eta(\para[x]))  \left( \delta(\perpe[x]-\pos) p^2 +\frac{1}{4}\delta''(\perpe[x]-\pos)\right)\right] } .
\end{align}
This expression can be readily expanded in powers of $\eta$.
Moreover, we expect that in a large variety of physical situations $\eta$ would be such that its average would vanish; 
as a consequence, we will  neglect the first order contribution, so that the first new contribution will appear at second order in $\eta$.
In formulae, we obtain
\begin{align}\label{eq:inhomo_HK2}
\begin{split}
 \Keta(x,y;T;\eta] &\approx \Kzeta(x,y;T)+ \Keta^{(2)}(x,y;T;\eta],
 \\
 \Keta^{(2)}(x,y;T;\eta]:&=\pathi[\mathcal{D}p \, ][x][0][y][T]e^{-\int_0^T\dx[t] \,\left[ p^2(t) -\mathi p(t)\dot{x}(t) - \ampli  \left( \delta(\perpe[x]-\pos) {\perpe[p]}^2 +\frac{1}{4}\delta''(\perpe[x]-\pos)\right)\right] }
 \\
 &\hspace{-1.cm}\times \int_0^T \int_0^{s_2}\, \dx[s_1] \dx[s_2] \,\prod_{j=1}^2\Bigg\{ 
 \eta\left(\para[x](s_j)\right) \left( {p_\dime}^2(s_j) +\frac{\partial^2_{L_j}}{4}\right)\delta\left(\perpe[x](_j)-\pos[j]\right)
 \Bigg\},
 \end{split}
\end{align}
in which we have explicitly used the symmetry under exchange of the intermediate times $s_{i=1,2}$.
Once more, the computation is more involved than the usual case, given that the potential on one side involves derivatives and  on the other should be regularized. 
For this reason, it proves convenient  to employ a series expansion in $\ampli$,  instead of trying to make direct use of the  heat-kernel in Eq.~\eqref{eq:HK_integral_analytic}.
Calling $\eta_j:=\eta\left(\para[x](s_j)\right)$ 
one obtains
\begin{align}\label{eq:inhomo_perturbative}
\begin{split} 
& \Keta^{(2)}(x,y;T;\eta]= \int_0^T \int_0^{s_2}\dx[s_1] \dx[s_2]\, 
   \sum_{n=2}^{\infty} {\ampli^{n}} 
\\
&\hspace{0.5cm}\times \int_0^T \dx[t_n]\cdots \int_0^{t_2} \dx[t_1]\,\prod_{j=1}^{n} \left( \delta_{\perpe[k](t_j)}^2 +\frac{\partial^2_{L_j}}{4}\right) \prod_{i=1}^{2} \left( \delta_{\perpe[k](s_i)}^2  +\frac{\partial^2_{L_i}}{4}\right)
 \\
& \hspace{0.5cm}\times  \pathi[][x][0][y][T]\,  
e^{\frac{1}{4}\int_0^T\dx[t] \,\left[ \mathi \dot{x}(t)+k(t)\right]^2 } \eta_1 \eta_2   \prod_{j=1}^{n} \delta\left(\perpe[x](t_j)-\pos[j]\right)\prod_{i=1}^{2}\delta\left(\perpe[x](s_i)-\pos[i]\right).
 \end{split}
\end{align}
If we now want to interpret the Dirac delta functions as fixing the path at given times, 
then there is an additional difficulty related to the fact that the intermediate times $s_{i=1,2}$ 
are not ordered with respect to the other times, $t_{i}$.
However, whatever value $s_{i=1,2}$ may take, 
they will of course fall into one of the intervals $(0,t_1)$, $(t_1,t_2)$, $\cdots$, $(t_n,T)$. 
We can thus define ordered times $\hat t_i(lm)$, such that 
\begin{align}
(\hat t_0(lm), \hat t_1(lm), \cdots):= (0, t_1, \cdots t_{l}, s_1, t_{l+1}, \cdots, t_{m}, s_2, t_{m+1},\cdots, T ),
\end{align}
and introduce an analogous definition for the\footnote{The hatted coordinates correspond to the $\dime$th component, i.e. $\hat x_i =\perpe[\hat x]_i$; we omit the upper index $\dime$ to simplify the notation.} $\hat L_i$ and $\hat x_i$. Hiding the $(lm)$ dependence for reasons of readability, we obtain
\begin{align}\label{eq:inhomo_perturbative15}
\begin{split} 
& \Keta^{(2)}(x,y;T;\eta]= 
  \frac{(-1)}{2^7 \pi^{3/2}}  \sum_{n=0}^{\infty}      \sum_{l\leq m=0}^n \pathi[][\para[x]][0][\para[y]][T][{\para[x]}]\, \eta_1 \eta_2 \,e^{-\frac{1}{4}\int_0^T\dx[t] \, {{\dot{\para[x]}}}^2 } 
\\
  &\times   \Bigg\lbrace { \left(\frac{\ampli }{8\sqrt{\pi}}\right)^{n-m} }   \int_0^{T} \dx[s_2] \int_{s_2}^{T}\dx[ t_{n}]\cdots \int_{s_2}^{ t_{m+2}} \dx[ t_{m+1}] 
\frac{ \Delta \hat x_{n+2} }{ \Delta\hat t_{n+2}^{3/2} } 
\\
&\hspace{4cm}\times\left[ \prod_{r=m+2}^{n+1} \frac{(2\Delta\hat t_r- \Delta\hat x_r^2)}{\Delta\hat t_r^{5/2}} \right] \left[\prod_{p=m+2}^{n+2} e^{-\frac{\Delta\hat  x_p^2}{4\Delta \hat t_p}} \right]\Bigg\rbrace
\\
  &\times \Bigg\lbrace { \left(\frac{\ampli }{8\sqrt{\pi}}\right)^{m-l}  \int_0^{s_2} \dx[s_1] \int_{s_1}^{ s_2} \dx[ t_m]\cdots \int_{s_1}^{ t_{l+2}} \dx[ t_{l+1}]} 
  \left[ \prod_{r=l+1}^{m+1} \frac{(2\Delta\hat t_r- \Delta\hat x_r^2)}{\Delta\hat t_r^{5/2}} \right]
  \\
  &\hspace{6cm}\times\left[\prod_{p=l+1}^{m+1} e^{-\frac{\Delta\hat  x_p^2}{4\Delta \hat t_p}} \right]\Bigg\rbrace
  \\
  & \times\left\lbrace \left(\frac{\ampli }{8\sqrt{\pi}}\right)^l  \int_0^{s_1}\dx[ t_l]\cdots \int_0^{ t_2} \dx[ t_1] 
\frac{ \Delta \hat x_0 }{ \Delta\hat t_0^{3/2} } \left[ \prod_{r=1}^{l} \frac{(2\Delta\hat t_r- \Delta\hat x_r^2)}{\Delta\hat t_r^{5/2}} \right] \left[\prod_{p=0}^{l} e^{-\frac{\Delta\hat  x_p^2}{4\Delta \hat t_p}} \right]\right\rbrace.
\end{split}
\end{align}
Notice that we may also reorder the series to get the compact expression
\begin{align}\label{eq:inhomo_perturbative2}
\begin{split} 
 \Keta^{(2)}(x,y;T;\eta]
&=  
  - \frac{1}{16}\int_0^{T} \dx[s_2] \int_0^{s_2} \dx[s_1] \pathi[][\para[x]][0][\para[y]][T][{\para[x]}]\, \eta_1 \eta_2 \,e^{-\frac{1}{4}\int_0^T\dx[t] \, {{\dot{\para[x]}}}^2 }
\\
& \hu\times  S_1(\perpe[\tilde y];T-s_2; \ampli) S_2(s_2-s_1; \ampli) S_1(-\perpe[\tilde x]; s_1; \ampli) ,
\end{split}
\end{align}
where the $S_1$ and $S_2$ functions are defined as follows:
\begin{align}
 \begin{split}
  S_1(x; T; \ampli) :&= \frac{1}{2\sqrt{\pi}}\sum_{n=0}^{\infty}    \left(\frac{\ampli }{8\sqrt{\pi}}\right)^{n}  \lim_{\Delta x_{0},\cdots\Delta x_{n-1} \to 0} \int_{0}^{T}\dx[ \alpha_{n}]\cdots \int_{0}^{ \alpha_{2}} \dx[ \alpha_{1}] 
\frac{  x e^{-\frac{   x^2}{4\Delta \alpha_n}}}{ \Delta \alpha_{n}^{3/2} }
\\
  &\hspace{2cm}\times 
 \left[ \prod_{r=0}^{n-1} \frac{(2\Delta  \alpha_r- \Delta x_r^2)}{\Delta \alpha_r^{5/2}} \right] \left[\prod_{p=0}^{n-1} e^{-\frac{\Delta   x_p^2}{4\Delta \alpha_p}} \right],
\\
  S_2( T; \ampli):&= \frac{1}{2\sqrt{\pi}} \sum_{n=0}^{\infty}    \left(\frac{\ampli }{8\sqrt{\pi}}\right)^{n}  \lim_{\Delta x_{0},\cdots\Delta x_{n} \to 0}  \int_{0}^{T}\!\dx[ \alpha_{n}]\cdots \int_{0}^{ \alpha_{2}}\! \dx[ \alpha_{1}] 
  \left[\prod_{p=0}^{n} e^{-\frac{\Delta   x_p^2}{4\Delta \alpha_p}} \right]\\
  &\hspace{2cm}\times 
  \left[ \prod_{r=0}^{n} \frac{(2\Delta  \alpha_r- \Delta x_r^2)}{\Delta \alpha_r^{5/2}} \right] .
 \end{split}
\end{align}
In the usual case, i.e when one considers potentials that do not involve derivatives, the $S_i$ functions would correspond to heat-kernels with constant coupling. 
In the current model, the $S_i$ play instead the role of regularized derivatives of the heat-kernel with constant coupling $\ampli$.
Employing the results in Apps. \ref{app:integral_hermite} and \ref{app:series_hermite},  we can perform a resummation of
the power series in $\ampli$  and obtain them in closed form:
\begin{align}
 \begin{split}
   S_1(x; T; \ampli) &= \frac{e^{-\frac{x^2}{4T}}}{2\sqrt{\pi} T^{}} \sum_{n=0}^{\infty}    \left(-\frac{\ampli  \,\text{sign}(x) }{4 \sqrt{T}}\right)^{n}    H_{n+1}\left(\frac{x}{2\sqrt{T}}\right)
   \\
   &= - \frac{4 }{\ampli^2  } \text{sign}(x) \left[{ e^{\frac{4 T+2 \ampli  |x|}{\ampli ^2}} \text{erfc}\left(\frac{\ampli  |x|+4 T}{2 \ampli  \sqrt{T}}\right)} - \ampli\frac{e^{-\frac{x^2}{4T}}}{\sqrt{4\pi T}} \right],
 \end{split}
 \\
 \begin{split}
   S_2( T; \ampli) &=- \frac{1}{2\sqrt{\pi} T^{3/2}} \sum_{n=0}^{\infty}    \left(-\frac{\ampli }{4\sqrt{T}}\right)^{n}    H_{n+2}\left(0 \right)
   \\
   &=  - \frac{16}{\zeta^3} \left[{ e^{\frac{4 T}{\ampli ^2}} \text{erfc}\left(\frac{4 \sqrt{T}}{2 \ampli  }\right)} -\frac{\ampli}{\sqrt{4\pi T}}  \right].
   \end{split}
\end{align}
At this point we can perform the following fast check. If we consider a constant $\eta$,
then the result in Eq. \eqref{eq:inhomo_perturbative2} reduces, as expected, to a free heat-kernel in the parallel directions, 
multiplied by $K_{\ampli+\eta}(x,y;T)$ restricted to quadratic order in $\eta$. 

Considering once more the expression \eqref{eq:inhomo_perturbative2} for the heat-kernel, the integrals in the parallel directions are trivial
\begin{align}
 \begin{split}
   \pathi[][\para[x]][0][\para[y]][T][{\para[x]}]\, \eta_1 \eta_2 \,e^{-\frac{1}{4}\int_0^T\dx[t] \, {{\dot{\para[x]}}}^2 }
&= \frac{e^{-\frac{(\para[y]-\para[x])^2}{4T}}}{(4\pi T)^{(\dime-1)/2}} \int \frac{\dx[{\para[k_1]}] \dx[{\para[k_2]}]}{(2\pi)^{2\dime-2}} \tilde\eta_1 \tilde\eta_2  
\\
&\hspace{-3cm} e^{\mathi (\para[k_1]\cdot \para[x_c](s_1)+\para[k_2]\cdot \para[x_c](s_2)) }  e^{-\frac{(T-s_1)s_1}{T} {\para[k_1]}^2-\frac{(T-s_2)s_2}{T} {\para[k_2]}^2-2\frac{s_1(T-s_2)}{T}\para[k_1]\cdot \para[k_2]} ,
 \end{split}
\end{align}
where  $\para[x_c](t):= \frac{(\para[y]-\para[x])}{T} t +\para[x]$ 
and the Fourier transforms $\tilde\eta_i:=\tilde \eta({\para[k_i]})$ are defined as
\begin{align}
  \eta(\para[x])=: \int \frac{\dx[{\para[k]}]}{(2\pi)^{\dime-1}} e^{\mathi \para[k] \cdot\para[x]} \tilde \eta({\para[k]}).
\end{align}
Replacing in Eq. \eqref{eq:inhomo_perturbative2} we obtain our final expression for the contribution to the heat-kernel of quadratic order in $\eta$:
\begin{align}\label{eq:master_inhomo}
\begin{split} 
 \Keta^{(2)}(x,y;T;\eta]
&= - \frac{1}{16}  \frac{e^{-\frac{(\para[y]-\para[x])^2}{4T}}}{(4\pi T)^{(\dime-1)/2}} \int_0^{T} \dx[s_2] \int_0^{s_2} \dx[s_1]\int \frac{\dx[{\para[k_1]}] \dx[{\para[k_2]}]}{(2\pi)^{2\dime-2}} \tilde\eta_1 \tilde\eta_2  
\\
&\hspace{-1cm}\times e^{\mathi (\para[k_1]\cdot \para[x_c](s_1)+\para[k_2]\cdot \para[x_c](s_2)) }  e^{-\frac{(T-s_1)s_1}{T} {\para[k_1]}^2-\frac{(T-s_2)s_2}{T} {\para[k_2]}^2-2\frac{s_1(T-s_2)}{T}\para[k_1]\cdot \para[k_2]}
   \\
&\hspace{-1cm}\times  S_1(\perpe[\tilde y];T-s_2; \ampli) S_2(s_2-s_1; \ampli) S_1(\perpe[\tilde x]; s_1; \ampli) 
.
\end{split}
\end{align}

\subsection{Mapping the heat-kernel: Neumann semitransparent to Dirichlet semitransparent}\label{sec:map}
Looking at the closed expressions for the $S_i$ one notices that they are proportional to the heat-kernel for a delta potential, cf. Eq. \eqref{eq:hk_delta}. 
The explicit relations are
\begin{align}
 S_1(x-\pos; T; \ampli)&= \frac{4\,\text{sign}(x-\pos)}{\ampli} \Kdelta(x,\pos ;T;4\ampli^{-1}),
 \\
 S_2(T; \ampli) &= \frac{16}{\ampli^2} \Kdelta(\pos,\pos ;T;4\ampli^{-1}).
\end{align}
A more carefully comparison shows that the proportionality extends also to the heat-kernel expression at quadratic order in $\eta$, 
so that, up to a rescaling in the inhomogeneities,  
the weak-background problem in the delta case (or more precisely, local semi-transparent Dirichlet boundary conditions) is mapped to a strong-background regime in our current model and \emph{vice versa}\footnote{We are defining $\Kdelta^{(2)}(x,y ;T;4\ampli^{-1};-4\ampli^{-2}\eta]$ as the contribution to the heat-kernel of the operator $A^{\rm delta}_{\dime}:=\left[-\partial^2+m^2+ \big( \eta(\para[x])+\ampli \big)  \delta(\perpe-\pos)\right]$ which is quadratic in $\eta$.}:
\begin{align}
  \begin{split}
\Kzeta^{(2)}(x,y;T;\eta] &= \operatorname{sign}( \perpe[\tilde x]) \operatorname{sign} (\perpe[\tilde y]) \Kdelta^{(2)}(x,y ;T;4\ampli^{-1};-4\ampli^{-2}\eta].
 \end{split}
\end{align}
In particular, this means that we may borrow some results from \cite{Franchino-Vinas:2020okl}; 
as an example, the trace of the heat-kernel (which will be employed to compute the effective action) is simply given by 
\begin{align}\label{eq:HKtrace2}
\begin{split} 
& \operatorname{Tr} \Keta^{(2)}(x,y;T;\eta]
=   \frac{8}{\ampli^4} \frac{T^2}{(4\pi T)^{(\dime-1)/2}} \int \frac{\dx[{\para[k_1]}] }{(2\pi)^{\dime-1}} |\tilde\eta_1|^2 \int_0^{1} \dx[s_-]  
    e^{- T \para[k_1]^2 s_- (1-s_-)} 
\\
&\hu \times \Kdelta(\pos,\pos ;T(1-s_-);4\ampli^{-1}) \Kdelta(\pos,\pos ;Ts_-;4\ampli^{-1}).
\end{split}
\end{align}

An even more detailed inspection shows that this is not an accidental relation valid only for the quadratic expansion in $\eta$. 
Indeed, one can repeat the computations performed in the previous section for an arbitrary order in $\eta$, 
to find that, dismissing the path integrals in the parallel directions, one obtains a chain 
\begin{align}
\begin{split}
\Keta^{(n)}(x,y;T;\eta]&\sim \int \dx[s_1]\cdots \dx[s_n]\, S_1( {\perpe[\tilde y]};T-s_n; \ampli) 
\\
&\hu\times S_2(s_n-s_{n-1}; \ampli) \cdots S_2(s_2-s_1; \ampli) S_1(-{\perpe[\tilde x]}; s_1; \ampli),
\end{split}
\end{align}
which is exactly the one we would have obtained in the delta case. 
This provides an order by order proof of the existence of a map
between the heat-kernels of these two different problems; 
more precisely, the map is given by
\begin{align}\label{eq:hk_duality}
  \begin{split}
\Kzeta(x,y;T;\eta]&
= \operatorname{sign}(\perpe[\tilde x]) \operatorname{sign} (\perpe[\tilde y]) \left[ \Kdelta(x,y ;T;4\ampli^{-1};4\ampli^{-2}\eta]-\frac{e^{-\frac{(x-y)^2}{4 T}}}{(4\pi T)^{\dime/2}} \right]
  \\
  &\hspace{-0.5cm}+\frac{\left(1+ \text{sign}(\perpe[\tilde x]) \text{sign}(\perpe[\tilde y]) \right) }{2 (4 \pi T )^{\dime/2}}  \left( e^{-\frac{(x-y)^2}{4 T }}+ e^{-\frac{(\perpe[\tilde x]+\perpe[\tilde y])^2+(\para[x]-\para[y])^2}{4 T }}\right).
 \end{split}
\end{align}
This map depends strongly on the fact that $\eta$ lives on the plate and will be broken if one introduces for example an additional potential with support outside from it.
\iffalse
To understand this point, notice that in such a case one would need to replace the function $S_2$ in the expansion of the heat-kernel with\footnote{Recall the notation introduced in Eq.~\eqref{eq:xtilde}.}
\begin{align}
\begin{split}
   S_3(x; T; \ampli) 
   &=  - \frac{16}{\zeta^3} \left[{ e^{\frac{4 T+2 \ampli  |\tildex|}{\ampli ^2}} \text{erfc}\left(\frac{\ampli  |\tildex|+4 T}{2 \ampli  \sqrt{T}}\right)} -\ampli\frac{e^{-\frac{\tildex^2}{4T}}}{\sqrt{4\pi T}} \left(1- \frac{\ampli |\tildex|}{4 T}\right) \right],
   \end{split}
\end{align}
multiply by the potential and integrate over $x$ in the whole space. 
The proportionality to the delta case is recovered only if we constraint $x=0$, where we have $S_2(T;\ampli)=S_3(0;T;\ampli)$. 
\fi
As we will see in Sec.~\ref{sec:ea}, 
the mapping that we have discussed will automatically translate into a duality at the level of the renormalized semiclassical field theory.

\subsection{The purely inhomogeneous coupling}
One interesting case is that in which the background field is small, such that the fluctuations may become larger than it\footnote{But keeping always $\ampli+\eta>0$ such to avoid instabilities triggered by possible bound states}. 
In this regime we will be able to obtain a closed expression for the trace of the  heat-kernel 
and the effective action in the massive  case. 
Moreover, as we will see this will turn out to be an instructive limiting case.

Let us then begin with expression~\eqref{eq:HKtrace2}. 
Taking its small-$\ampli$ limit involves expanding the heat-kernel of the delta potential for large coupling, which gives
\begin{align}
\begin{split}
&\Kdelta(x,x';T;4\ampli^{-1})=
\frac{e^{-\frac{(x-x')^2}{4 T }}}{2 \sqrt{\pi } \sqrt{T }}-\frac{e^{-\frac{(\left| \tildex \right| +\left| \tildey\right| )^2}{4 T }}}{2 \sqrt{\pi } \sqrt{T }}
+\frac{\zeta  (\left| \tildex\right| +\left| \tildey\right| ) e^{-\frac{(\left| \tildex\right| +\left| \tildey\right| )^2}{4 T }}}{8 \sqrt{\pi } T ^{3/2}}
\\
&\hu
-\frac{\zeta ^2 e^{-\frac{(\left| \tildex\right| +\left| \tildey \right| )^2}{4 T }} \left(\left( \left| \tildex \right| + \left| \tildey \right|\right)^2-2 T \right)}{32 \sqrt{\pi } T ^{5/2}}+\mathcal{O}(\ampli^3),
\end{split}
\end{align}
so that after setting $x'\equiv \pos$ the leading term cancels, rendering the expression more divergent as $T\to 0$:
\begin{align}
 \Kdelta(x,\pos ;T;4\ampli^{-1})&=
 \frac{\zeta  e^{-\frac{(L-x)^2}{4 T }} (2 \left| L-x\right| +\zeta )}{16 \sqrt{\pi } T ^{3/2}}-\frac{\zeta ^2 (L-x)^2 e^{-\frac{(L-x)^2}{4 T }}}{32 \sqrt{\pi } T ^{5/2}}+\mathcal{O}(\ampli^3).
\end{align}
Setting also $x\equiv \pos$, the first contribution to $\Kdelta$ is of order $\ampli^2$, 
which is exactly the power needed to cancel the inverse powers of $\ampli$ in Eq.~\eqref{eq:HKtrace2}.
In our computation the limit $x,x'\to \pos$ should be taken at the end, since they act as regulators;  the final expression once the mass contribution is reinstated is given by
\begin{align}\label{eq:HKtrace_pure}
\begin{split} 
& \operatorname{Tr} \Keta[0]^{(2)}(x,y;T;\eta]
= -\frac{1}{2^5 }\frac{e^{-m^2T}}{T(4\pi T)^{(\dime-1)/2} } 
   \int \frac{\dx[{\para[k]}] }{(2\pi)^{\dime-1}} |\tilde\eta|^2 \,g\!~\left(T\para[k]^2\right),
\end{split}
\end{align}
where  $g(\cdot)$ is defined in terms of the modified Bessel functions of the first kind $I_\alpha(\cdot)$:
\begin{align}
g(b):&= b e^{-\frac{b}{8}} \left[I_0\left(\frac{b}{8}\right)+I_1\left(\frac{b}{8}\right)\right].
\end{align}

To have an intuition of the result~\eqref{eq:HKtrace_pure}, one can study the behaviour of $g(\cdot)$ for large and small arguments:
\begin{align}
 &g(b)
 =\begin{cases}
  \frac{4 \sqrt{b}}{\sqrt{\pi }}-\frac{4 }{\sqrt{\pi b}}-\frac{6 }{\sqrt{\pi b^3}} + \mathcal{O}\left(b^{-5/2}\right),
 \\
b-\frac{b^2}{16}+\frac{b^3}{256}+\mathcal{O}\left(b^4\right).
 \end{cases} 
\end{align}
Notice that the limit of a homogeneous configuration gives a vanishing contribution, 
in agreement with the previous results in Sec. \ref{sec:worldline_hk}. 
Moreover, the first terms in a small-propertime expansion are local.
If instead the field $\eta$ acquires modes with large momenta, then the expansion involves half-integer powers
of $\para[k]$, which is tantamount of saying that nonlocal terms will play an important role.

The result in \eqref{eq:HKtrace_pure} deserves one last comment,
which will become important in the analysis of the effective action in the following section. 
A consequence of the small $\ampli$ limit is that the expression for the heat-kernel, in the expansion for small propertime, is more divergent than the finite $\ampli$ case by a factor $T^{-1}$.
More generally, every time we increase the order in $\eta$ by one, the expansion of the heat-kernel for small $T$ will be more divergent by a factor $T^{-1/2}$.
This is reminiscent of the findings for a constant coupling, cf. \eqref{eq:hk_zeta_n}, 
signaling that for a small coupling a resummation may be needed in order to see the real behaviour for small $T$.

\section{The effective action}\label{sec:ea}

\subsection{Duality for the  Field Theory}\label{sec:QFT_duality}
Consider now the mapping in Eq.~\eqref{eq:hk_duality} at the level of the effective action
adding a mass term. 
Employing the formula~\eqref{eq:EA_trlog} for the effective action in terms of the heat-kernel's trace,
we get the following relation between the 
quantum contributions to the effective action in our generalized Neumann case, $\Gamma_{\rm 1-loop}$, 
and those in the generalized Dirichlet case, $\Gamma_{\rm 1-loop}^{\rm delta}$: 
\begin{align}\label{eq:EA_duality}
  \begin{split}
\Gamma_{\rm 1-loop}(\ampli;\eta]&=
-\frac{1}{2}\int_0^\infty \frac{\dx[T]}{T}\, e^{-m^2 T} \int \dxd[x][\dime] \, \Kzeta(x,x;T;\eta] 
\\
&\hspace{-1cm}= -\frac{1}{2}\int_0^\infty \frac{\dx[T]}{T}\,  e^{-m^2 T}\int \dxd[x][\dime] \,  \left[ \Kdelta\left(x,x ;T;4\ampli^{-1};4\ampli^{-2}\eta\right]+\frac{ e^{-\frac{(x-\pos)^2}{ T }} }{(4\pi T)^{\dime/2}}\right]
\\
 &\hspace{-1cm}=\Gamma^{\rm delta}_{\rm 1-loop}\left(4\ampli^{-1};4\ampli^{-2} \eta\right]+4^{-\dime/2}.
 \end{split}
\end{align}
Inasmuch as we don't consider the interaction with gravity, the constant factor $4^{-\dime/2}$ is irrelevant in the computation of physical quantities, since it will be absorbed in the renormalization of the cosmological constant. 
 Therefore, we can see that there is a duality between  both theories at the quantum level:
if one desires to compute the large background expansion in one theory, one may simply study the small background of the other.
These assertions are valid independently from the dimensions $\dime$ of spacetime in which we choose to work.

\subsection{The inhomogeneous and massless case in \texorpdfstring{$\dime=4$}{}}\label{sec:ea_inho}
As an immediate consequence of the duality discussed in the previous paragraphs,
we can study the massless case in $\dime=4$ by borrowing  results 
from the delta-potential case previously obtained by some of the authors of this work~\cite{Franchino-Vinas:2020okl}.
The explicit result for the effective action at quadratic order in $\eta$ is
\begin{align}\label{eq:ea_inho}
\begin{split}
 \Gamma^{(2)}_{m=0,\,\dime=4}:&=\frac{1}{2} \text{Tr}\,\text{Log} \,\opeinho \,\Big\vert_{\text{order }\eta^2}\\
 &=\int \frac{\dxd[k^{\parallel}][3]}{(2\pi)^{3}} \, \tilde \eta(\para[k]) \tilde \eta(-\para[k]) \ff(\para[k],\ampli),
 \end{split}
\end{align}
where the form factor $\ff$ has been split into three terms, one divergent as $\dime$ tends to four, 
another which is finite and local ($\ff_L$), and the remaining which is finite and nonlocal ($\ff_{NL}$):
\begin{align}\label{eq:ff}
 \ff(\para[k],\ampli):&=-\frac{8}{\ampli^4}\left[\frac{1}{4\pi^2 \ampli}\frac{1}{D-4}+\ff_{L}(4\ampli^{-1})+\ff_{NL}(\para[k],4\ampli^{-1})\right].
\end{align}
The explicit expressions for these contributions, defining $b^2:=16({\ampli k})^{-2} $ and in terms of Lerch's transcendent function $\Phi(\cdot,\cdot,\cdot)$, read
\begin{align}
 \ff_{L}(4\ampli^{-1}):&=\frac{1}{8 \pi ^2\ampli } \left(\gamma -4+\log \left(\frac{\mu^2\ampli^2}{256 \pi ^3}\right)\right),\label{eq:ffl}\\
\ff_{NL}(k,4\ampli^{-1}):&=\frac{1}{8\pi^2\ampli}\left[-\frac{1}{b}
+ \left(2+b\right) \log\left(1+\frac{2}{b} \right)+\frac{ b }{4(1+b)} \Phi\left(\frac{1}{(1+b)^2},2,\frac{1}{2} \right)
\right]\,
.\label{eq:ffnl}
\end{align}
 In order to render the nonlocality of $\ff_{NL}$ more visible, one can perform expansions for large and small $b$, 
 for which we get either $\log (k^2)$ contributions or half-integer powers of $k^2$
 that preclude a so-called derivative expansion (see~\cite{Fosco:2011xx} for its application to a Casimir configuration which is similar in spirit to ours):
 \begin{align}\label{eq:fnl_exp}
  \ff_{NL}(k,4\ampli^{-1})&=
  \begin{cases}
-\frac{k}{32\pi^2 }\left[1-4 \log \left(\frac{\ampli^2  k^2}{4}\right)-\frac{2 \left(4 \log (\ampli^2  k^2/4)+\pi ^2+8\right)}{\ampli  k}+\mathcal{O}\left(({k \ampli})^{-2}\right)\right],
  \\
  \frac{3}{8 \pi ^2 \ampli}\left[1-\frac{\ampli ^2 k^2}{216}+\frac{\ampli ^4 k^4}{7200}-\frac{\ampli ^5 (k^2)^{5/2}}{17280}\right.
  \\
  \hspace{3cm}\left.+\frac{11 \ampli ^6 k^6}{564480}-\frac{\ampli ^7 (k^2)^{7/2}}{161280}+\mathcal{O}\left((k \ampli)^8\right)\right].
 \end{cases}
 \end{align}

At this point some comments are in order. First,
 the leading terms were to be expected  from a simple dimensional analysis of the problem. 
 Indeed, this is the reason why corrections proportional to $k$ are so frequently encountered in the bibliography~\cite{MaiaNeto:2005ubc}.
 
 Second,
 the vanishing Neumann limit of $\Gamma^{(2)}$ seems to be well-motivated: indeed, a small variation around infinity should make no difference, 
 at least as long as $\eta$ is small, which was one of our hypothesis. The physical mechanism is similar to that in the Dirichlet case, where a larger coupling tends to repel the quantum field from the sheet, 
correspondingly attenuating the interaction with the background $\eta$.
The only difference is that the coupling in the current situation involves derivatives of the quantum field.

 Third,  Eq.~\eqref{eq:ff} signals that the theory needs to undergo renormalization. 
 In dimensional regularization, the only term that needs a counterterm is the mass term for $\eta$. 
 However, in other schemes one may obtain additional divergent terms, as discussed for general cases in \cite{Beneventano:1995fh, Kirsten:2001wz}.
 Although this is not the case if $\ampli>0$, the $\ampli=0$ case   is more subtle and will be discussed in Sec. \ref{sec:ea_pure}.

 Lastly  and related to the previous point, in the limit of vanishing  coupling the effective action  in expression \eqref{eq:ea_inho} is divergent, as can be seen from Eq.~\eqref{eq:ff} together with the corresponding definitions and the expansion in Eq.~\eqref{eq:fnl_exp}. 
 Such divergence simply indicates the fact that the expansion for small $\eta$ and $\ampli$ do not commute. 
 To gain insight into this point, consider the constant coupling case. 
 A straightforward computation shows that
 \begin{align}
 \frac{\Gamma^{}_{}}{V_{\parallel}}=\frac{1}{3 \pi ^3 (\dime-4) \ampli^3}+\frac{- \log (\pi \ampli^2 \mu^2 )-\psi ^{}\left(\frac{5}{2}\right)}{6 \pi ^3 \ampli ^3}+ \mathcal{O}(\dime-4),
\end{align}
where $\psi(\cdot)$ is the polygamma function and $V_{\parallel}$ is the volume over the plate. 
If we now consider homogeneous perturbations by replacing $\ampli\to \ampli +\eta$, the expansion will be in powers of $\eta/\ampli$, rendering clear our statement.
The physical intuition of why the expansion is singular for $\ampli$ around zero is related to the instabilities generated by the bound state for $\ampli<0$.
An alternative heuristic way to make sense of those divergences is to interpret them
as a need  of an additional renormalization. 
We will analyze this point further in the following section.

\subsection{On the purely inhomogeneous and massive case}\label{sec:ea_pure}
To understand better the $\ampli\to 0$ limit of the previous expressions, 
let us set $\ampli\equiv0$ right from the beginning;
the simpler formulae enable us to include the effects of a nonvanishing mass.
One can then compute the effective action employing the heat-kernel's trace in Eq. \eqref{eq:HKtrace_pure};
the result for a massive field in such case is\footnote{We are considering $\eta(\para[x])>0$, 
so that effectively one may extract a mean value and perturbations around it. 
However, it proves convenient for the following discussion to keep $\eta$ as one single entity.}
\begin{align}\label{eq:ea_pure}
 \Gamma^{(2)}_{\rm pure}
 &= \int \frac{\dxd[k^{\parallel}][3]}{       (2\pi)^{3}} \, \tilde \eta(\para[k]) \tilde \eta(-\para[k]) \ff_{\rm pure}(\para[k],m),
\end{align}
where the form factor is defined in terms of the hypergeometric function ${}_2F_1(\cdot,\cdot;\cdot;\cdot)$:
\begin{align}\label{eq:ff_pure}
 \ff_{\rm pure}(k,m):&=\frac{k^2 \left(m^2\right)^{3/2}}{384\pi} \,{}_2F_1\left(-\frac{3}{2},\frac{1}{2};2;-\frac{k^2}{4 m^2}\right) .
\end{align}
The asymptotic expressions of this form factor in the limits of large and small mass can be obtained from the corresponding expansions of the hypergeometric function, to read 
\begin{align}
  _2F_1\left(-\frac{3}{2},\frac{1}{2};2;-b\right)
 &=\begin{cases}
  \frac{8b^{3/2}}{15 \pi }+\frac{2\sqrt{b}}{\pi }+\mathcal{O}\left(b^{-1/2}\right),
    \\
   1+\frac{3 b}{8}+\frac{3 b^2}{64}-\frac{5 b^3}{1024}+\mathcal{O}\left(b^4\right).
   \end{cases}
\end{align}
In particular, the massless limit corresponds to a nonlocal term proportional to a half-integer power of $k^2$, namely $(k^2)^{5/2}$.

Notice that the result in Eq.~\eqref{eq:ea_pure} is automatically finite in dimensional regularization. 
However, as mentioned in the previous section the situation may change in other schemes. 
If instead of dimensional regularization a cutoff is introduced, 
then also the terms $(k^2)^i |\tilde\eta|^2$,  $i=1,2$, should be renormalized. 
Indeed, computing the effective action from \eqref{eq:HKtrace_pure},  
we introduce a UV cutoff $\Lambda$ with dimensions of momentum, set $\dime\equiv 4$ and expand for small $T$ to obtain
\begin{align}\label{eq:pure_divergent}
 \begin{split}
\Gamma^{(2)}_{\rm pure}&= \int \dxd[{\para[k]}][3]\int_{\Lambda^{-2}}^1 \dx[T] \,\vert \tilde \eta \vert^2 
\Bigg[ \frac{{\para[k]}^2}{512 \pi ^{3/2} T^{5/2}}-\frac{{\para[k]}^4+16{ \para[k]}^2 m^2}{8192 \pi ^{3/2} T^{3/2}}+\mathcal{O}\left(T^{-1/2}\right)\Bigg]+\cdots
 \\
 &= \frac{1}{768  \pi ^{3/2}}\int \dxd[{\para[k]}][3] \,\vert \tilde \eta \vert^2  \Bigg[ {{\para[k]^2} \Lambda   \left(\Lambda ^2-3 m^2\right)}-\frac{{\para[k]}^4 \Lambda }{16} \Bigg]+\cdots,
 \end{split}
\end{align}
where the dot points denote finite terms as $\Lambda$ tends to infinity. 
As a consequence, the theory lacks predictivity for the terms depicted in~\eqref{eq:pure_divergent}.
In particular, one may set all of them to zero, as in the substraction of large mass terms suggested in \cite{Bordag:1998vs}.

At this point one may understand the $\ampli \to 0$ limit of the results in the previous section as follows. 
The divergent contributions in Eq.~\eqref{eq:ea_pure} as $\ampli\to 0$ should be reabsorbed in a renormalization process;
the explicit equivalence between both approaches can be seen by comparing them to Eq.~\eqref{eq:pure_divergent}.
Taking this comment into account, one then sees that the $\ampli\to 0$ limit of Eq.~\eqref{eq:ea_inho} and the 
massless limit of expression~\eqref{eq:ea_pure} agree at the renormalized level.

One can also envisage what would happen once higher powers in $\eta$ are considered. 
At the end of Sec. \ref{sec:inhomogeneous} we have mentioned that, in the purely inhomogeneous scenario, 
the expansion of the heat-kernel for small propertimes acquires one additional $T^{-1/2}$ for every extra power of $\eta$.
This implies that the number of terms to be renormalized will also correspondingly increase. 
However, this situation is reminiscent of the perturbative expansion of the  heat-kernel for homogeneous coupling $\ampli$, 
cf. Eq.~\eqref{eq:hk_zeta_n}, where arbitrary large negative powers of the propertime appear. 
Once the series is resummed, the result \eqref{eq:HK_integral_analytic}  is seen to have only a $T^{-1/2}$ divergence for small propertime.
We expect that a similar mechanism should be behind the need for renormalization of terms like those in \eqref{eq:pure_divergent} as long as $\eta$ is strictly bigger than zero, 
since we have not seen them in the expansion studied in Sec. \ref{sec:ea_inho}. 
In other words, the resummation in $\ampli$ performed in Sec. \ref{sec:inhomogeneous}, if $\eta/\ampli<1$, is expected to be enough to avoid the singularities of the effective action's expansion for vanishing total coupling ($\ampli+\eta$).

\section{Dynamical Casimir effect and particle creation}\label{sec:dynamical}
Up to this point we have restricted ourselves to the consideration of the theory in Euclidean space. 
As customarily done, one can appeal to a Wick rotation in order to consider the problem in Minkowski space. This will allow us to mimic a situation of dynamical Casimir effect
through time-dependent properties of the wall; 
for more information on this effect, see  
the reviews~\cite{Dodonov:2020eto, Dodonov:2010zza} and references therein.

To examine this dynamical scenario, we will assume that the argument of the delta function in Eq.~\eqref{eq:action} corresponds to a spatial coordinate,
so that time may only be an argument of the external field $\eta$.
Calling $\tau$ the Euclidean time and $x_0$ the Minkowski one, the rotation $x_0=:-\mathi \tau$ 
(accompanied by analogous rotations for every $0$-component of a tensor) may be performed without encountering singularities in the form factors studied in this manuscript.
 One obtains then a master formula for the  effective action $\Gamma_M^{(2)}$ in Minkowski space  at second order in~$\eta$,
\begin{align}\label{eq:ea_minkowski}
 \Gamma_M^{(2)}&= -\int \frac{\dxd[\para[k]][3]}{(2\pi)^{3}} \, \left\vert \tilde \eta_M ( k_0,\paras[k])\right\vert^2  \ff\left(\sqrt{{\paras[k]}^2-k_0^2-\mathi \epsilon},m\right),
\end{align}
where we have introduced Feynman's prescription through an infinitesimal parameter $\epsilon$, we have taken the branch cut of the square root to be in the negative real axis and  the Fourier transform 
in Minkowski space is defined as\footnote{The set of components that are parallel to the plates and space-like is denoted by $\paras[k]$. }
\begin{align}
 \eta(\para[x])=:\int \frac{\dx[k_0]\dx[{\paras[k]}]}{(2\pi)^3} e^{\mathi (-x_0 k_0+ \paras[x]\cdot \paras[k])} \tilde\eta_M(k_0,\paras[k]).
\end{align}
Of course the form factor $F$, depending on the situation under study, may be chosen among those in Eqs.~\eqref{eq:ff} or~\eqref{eq:ff_pure}.

The expression for the effective action may be used to compute the creation of particles in a dynamical situation. 
Indeed, in the usual in-out formalism, the vacuum persistence's amplitude is given by the effective action as
\begin{align}
 \langle 0_{\rm out }\vert 0_{\rm in}\rangle   =e^{\mathi \Gamma_M}.
\end{align}
If the effective action develops an imaginary contribution, 
which may be possible by the appearence of branch cuts in \eqref{eq:ea_minkowski} after the Wick rotation,
then the vacuum becomes unstable through a process of pair creation, 
whose probability $P$ is defined as
\begin{align}
 1-P:= e^{-2\operatorname{Im} \Gamma_M}.
\end{align}
In the most frequently studied situation, i.e. for weak pair-creation processes, we  may approximate $P\approx 2 \operatorname{Im} \Gamma_M$.
Therefore, the quantity in which we are interested is
\begin{align}
 P\approx2 \int \frac{\dxd[{\para[k]}][3] }{(2\pi)^{3} }
 \, \left\vert \tilde \eta_M( k_0,\paras[k])\right\vert^2  \operatorname{Im} \left[ \ff_{}\left(\sqrt{{\paras[k]}^2-k_0^2-\mathi \epsilon},m\right)\right].
\end{align}
Turning back to Eq.~\eqref{eq:ff_pure}, the hypergeometric function $_2F_1\left(-\frac{3}{2},\frac{1}{2};2;x\right)$ possesses a branch cut,
which in the case of the principal branch runs from  $1$ to $\infty$ on the real $x$-axis. We may thus recast the probability of pair creation as
\begin{align}\label{eq:probability}
 P=2 \int \frac{\dxd[\para[k]][3]}{(2\pi)^{3}} \, \left\vert \tilde \eta_M( k_0,\paras[k])\right\vert^2 \Theta(k_0^2-{\paras[k]}^2-4m^2) \operatorname{Im} \left[ \ff_{}\left(\sqrt{{\paras[k]}^2-k_0^2-\mathi \epsilon},m\right) \right],
\end{align}
where $\Theta(\cdot)$ is the Heaviside function that signals the threshold of the pair-creation process: 
the external field $\eta$ must provide at least the rest energy of two particles for the process to take place.
In the case described in Eq.~\eqref{eq:ea_inho}, 
such threshold is absent because particles are taken as massless.

\subsection{Harmonic perturbations}
A simple model that mimics the dynamical Ca\-si\-mir effect, introduced in \cite{Silva:2011fq} for the one-dimensional case and studied also in \cite{Franchino-Vinas:2020okl} for an inhomogeneous delta potential, is given by perturbations that are 
harmonic in time with frequency $\omega_0$; for simplicity we will consider it independent of the spatial coordinates\footnote{Following the discussion in Sec.~\ref{sec:ea_pure}, it should be clear that formally it is not enough to consider the amplitude $\eta_0$ small;
one should also think that there is an additional background $\ampli$ (not necessarily much) bigger than $\eta_0$ 
for the expansion in $\eta$ to be well-defined.
}:
\begin{align}\label{eq:eta_harmonic}
 \eta_H(t):&=\eta_0 \cos\left(\omega_0 t\right) e^{-\frac{\vert t\vert }{T}}, \quad \omega_0,T>0,\quad\eta_0\in\mathbb{R}.
\end{align}
In Eq.~\eqref{eq:eta_harmonic}, the exponential factor is employed to impose a boundary in time, 
since otherwise the number of pairs created becomes infinite. 
In the limit of large $T$, a straightforward computation shows that its Fourier transform satisfies
\begin{align}
\vert \eta_H(k^0)\vert^2&=\frac{\pi}{2} \eta_0^2 T \left[\delta(k^0-\omega_0)+\delta(k^0+\omega_0)\right],\quad  \omega_0T\gg1,
\end{align}
which, defining the threshold frequency $\omega_c:=\omega_0/(2m)$, thus leads to 
the following probability of pair creation rate per unit area of the plate
in the purely inhomogeneous case ($A$ denotes the area of the plate):
\begin{align}
\begin{split}\label{eq:pp_h}
 \frac{P_{H}}{A T}&= \eta_0^2  \, \Theta(\omega_0^2-4m^2) \operatorname{Im} \left[ \ff_{\rm pure}\left(-\mathi \omega_0+\epsilon,m\right) \right]
 \\
 &=\frac{\eta_0^2}{3072 \pi }\Theta \left(\omega _c^2-1\right) (m^2)^{3/2}\omega _0^2 \left({\omega _c^2}-1\right){}^3 \, _2F_1\left(\frac{3}{2},\frac{7}{2};4;1-{\omega _c^2}\right).
 \end{split}
\end{align}
 To derive the last line we have employed the result in \cite{NIST:2010} for the jump across the branch cut of the hypergeometric function,
 which is proportional to the desired imaginary part.

\begin{figure}[h!]
\begin{center}
 \begin{minipage}{0.46\textwidth}
 \vspace{0.0cm}\includegraphics[width=1.0\textwidth]{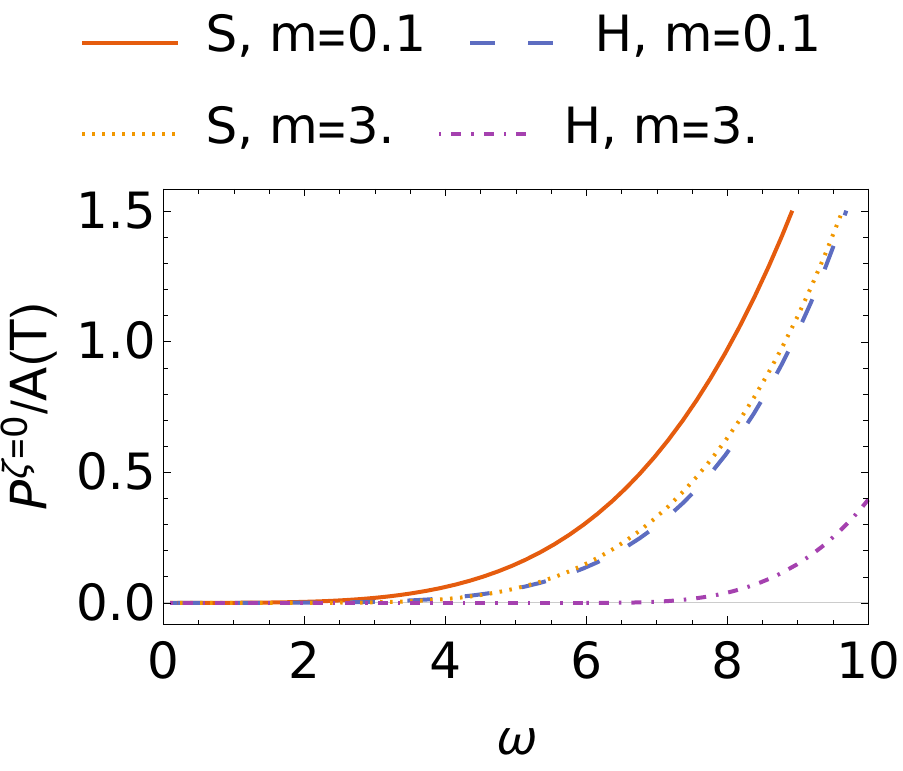} 
 \end{minipage}
\hspace{0.02\textwidth}
\begin{minipage}{0.46\textwidth}
 \vspace{0.0cm}\includegraphics[width=1.0\textwidth]{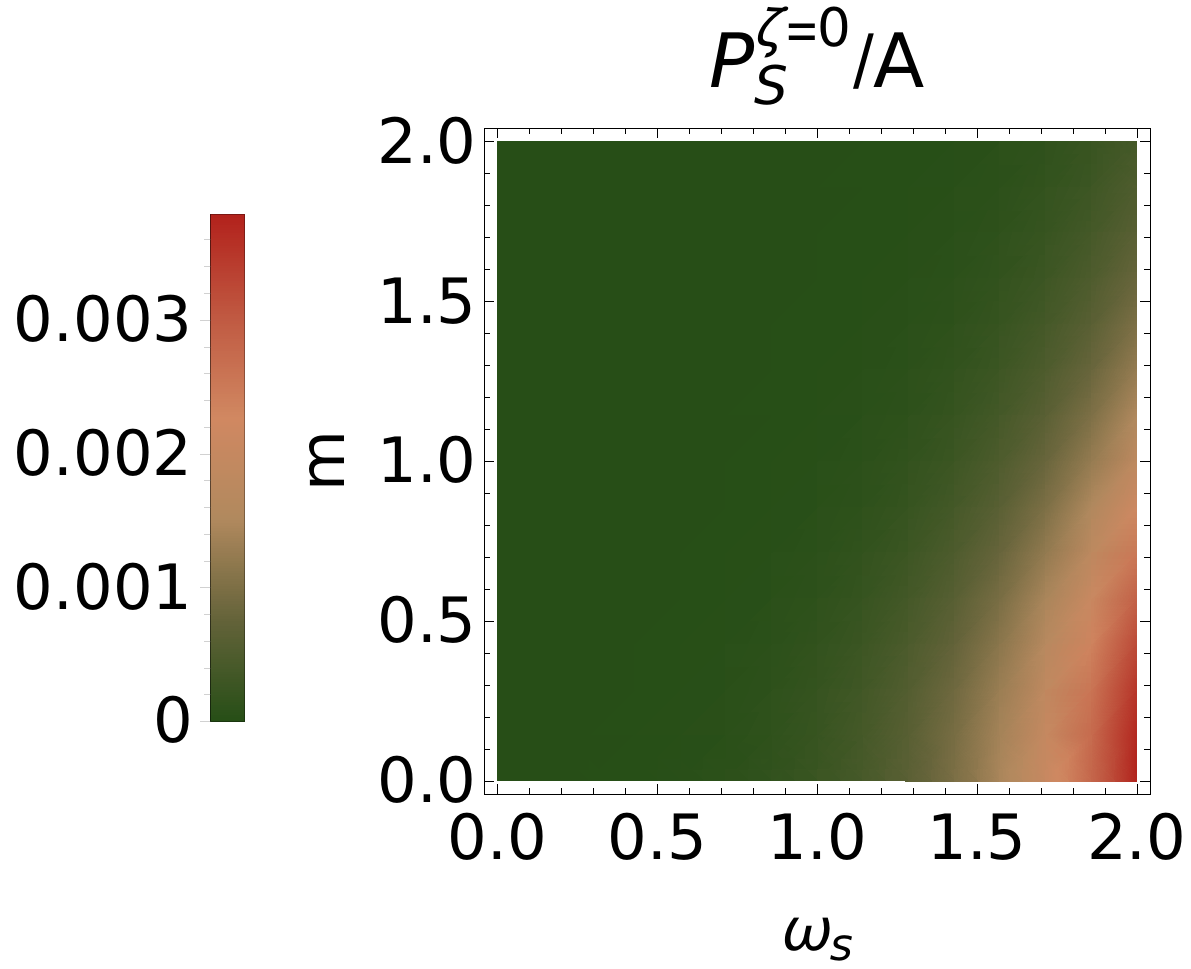} 
 \end{minipage}
 \end{center}
 \label{fig:massive}
 \caption{Pair production probability/rate per unit area in the massive case. 
 The left panel compares the harmonic and Sauter cases (denoted as $H$ and $S$ respectively) for different values of masses and as a function of the corresponding frequency. 
 Notice that in the harmonic case, the pair production rate should be understood.
 The panel on the right is a density plot of the pair production probability per unit area for the Sauter pulse as a function of the frequency $\omega_S$ and the mass $m$.
 In all the cases, the amplitudes $\eta_0$ are set to unity.}
 \end{figure}

 Analytically, we can compute the expansions for small and large $\omega_c$,
 which show that $\omega_0^5$ sets the scale of the probability rate:
\begin{align}\label{eq:pair_harmonic_massive}
\frac{P_{H}}{A T}&= \frac{\eta_0^2 \omega_0^5}{384\pi} \Theta(\omega _c^2-1)
\begin{cases}
&\displaystyle \frac{1}{15 \pi }-\frac{1}{4 \pi  \omega _c^2}+\frac{3 \left(4 \log \left(4\omega _c\right)-1\right)}{64 \pi  \omega _c^4}
\\
&\hu\displaystyle +\frac{-12 \log \left(4\omega _c\right)-5}{768 \pi  \omega _c^6}+\mathcal{O}\left({\omega _c}^{-7}\right),
\\\\
&\displaystyle \frac{1}{8} \left(\omega _c-1\right){}^3-\frac{33}{64} \left(\omega _c-1\right){}^4+\frac{345}{256} \left(\omega _c-1\right){}^5
\\
&\hu \displaystyle -\frac{2899 \left(\omega _c-1\right){}^6}{1024}+\mathcal{O}\left(\left(\omega _c-1\right){}^7\right).
\end{cases}
\end{align}
On one side, we see that in the massless limit we indeed recover (up to a rescaling) the infinite coupling result for the delta case~\cite{Franchino-Vinas:2020okl},
being the first corrections of order $\omega_c^{-2}$. Once more, on dimensional grounds this situation is reproduced in some analogous setups, such as a moving mirror~\cite{Fosco:2007nz}.

On the other side, for large masses the result evidently vanishes as a consequence of the mass threshold,
since the energy of the oscillations are not enough to provide the minimum energy of two particles at rest. 
For values of $\omega_c$ slightly larger than the threshold, the particle creation rate behaves as a third power
in the difference ($\omega_c-1$).
These behaviours can be confirmed from the plot in the left panel of Fig.~\ref{fig:massive}, 
where the pair production rate per unit area is shown as a function of the frequency; the dashed blue line and the violet dashed-dotted line correspond respectively to $m=0.1$ and $m=3$.

As a generalization of this simple harmonic example, one can also consider a perturbation that
resembles a plane wave over the plate,
\begin{align}
 \eta_{W}(x,t):&=\eta_0 \cos\big(k(x-vt)\big) e^{-{\sigma \vert x-vt\vert }}, \quad k,\sigma>0,\quad \eta_0,v\in\mathbb{R},
\end{align}
where $k$ is its wavenumber, $\sigma\to 0$ is a regulator and $v$ the speed of the wave 
(recall that $v=1$ in our units equals $c$, the speed of light in vacuum).
 After removing the regulator we get the following pair production rate per unit area:
 \begin{align}\label{eq:pp_wave}
\begin{split}
 \frac{P_{W}}{A T}&= \eta_0^2  \, \Theta\Big((v^2-1)k^2-4m^2\Big) \operatorname{Im} \left[ \ff_{\rm pure}\left(\sqrt{ (1-v^2)k^2  -\mathi \epsilon},m\right) \right].
 \end{split}
\end{align}
 On one side we see that pair production is possible only if $v\geq 1$.
 Notice that this situation should be understood not as a travelling wave with speed faster than light, 
 but rather as an active fast modulation of a property over the plate, 
 in a way analogous to that proposed for example in~\cite{Oue:2021ioa}.
 
 On the other side, defining an effective frequency $\omega^2_{\rm eff}:= (v^2-1)k^2$,
 the result~\eqref{eq:pp_wave} can be obtained from~\eqref{eq:pp_h}
 by simply trading $\omega_0\to \omega_{\rm eff}$.
 In particular, the exclusively time-dependent case $\eta_H$ can be understood as 
 its infinite speed limit (after an appropriate rescaling of $k$).

 \subsection{Sauter pulse }
One widely diffused profile in the literature of QED in external backgrounds is the Sauter pulse~\cite{Sauter:1932gsa},
defined as
\begin{align}
 \eta_{S}(t):=\frac{\eta_0}{\cosh^2(\omega_S t)}, \quad \omega_S\in\mathbb{R}-\{0\},\,\eta_0>0.
\end{align}
Replacing this profile in our general equation~\eqref{eq:probability},
we can compute the probability of pair creation $P_S$.

First, for the massive case (and $\ampli=0$), we show a density plot of $P^{\ampli=0}_S$ per unit area 
in the right panel of Fig.~\ref{fig:massive} as a function of the frequency $\omega_S$ and the mass $m$.
On the one hand, as could be expected, as the mass increases the probability of pair creation diminishes, 
since the cost of creating the pair becomes higher.
On the other, if the frequency $\omega_S$ becomes larger, the distribution of $\eta$ in Fourier space is widened,
so that creation of pairs is more favoured.
This is in agreement with the following analytical asymptotics 
\begin{align}\label{eq:pair_sauter_massive}
 \frac{P_S^{\ampli=0}}{ A}=
 \eta_0^2 \omega _S^4 \times \begin{cases}
  \frac{7 \zeta_{\rm R} (7) }{ \pi ^9} +\cdots,\quad m/\omega_S\ll 1 ,
  \\
 \frac{ 1}{4\pi ^4} \frac{m^4}{\omega _S^4}e^{-\frac{2 \pi  m}{\omega _S}} \left(1+\frac{23  }{4 \pi } \frac{\omega _S }{m}\right) +\cdots,\quad m/\omega_S\gg 1,
 \end{cases}
\end{align}
where $\zeta_{\rm R}(\cdot)$ is Riemann's zeta function. 
Although the exponential suppression for large masses may remind the one present in the Schwinger pair production for rather general electric fields~\cite{Dunne:2006st}, 
keep in mind that our process is perturbative in the background field amplitude and therefore intrinsecally different in nature.

Additionally, in the left panel of Fig.~\ref{fig:massive} we show the behaviour of the pair production probability per unit area as a function of the frequency $\omega_S$;
the solid red line and the yellow small-dashed line correspond respectively to $m=0.1$ and $m=3$.
Compared to the harmonic case, if frequencies are small we see that the exponential supression allows for a faster setting in of pair production; 
in other words, the Sauter pulse always embodies some frequency components above the mass threshold that enables the creation of pairs.
On the contrary, for larger frequencies the trend will be reverted and the harmonic pair production will become greater, as dictated by Eqs.~\eqref{eq:pair_harmonic_massive} and \eqref{eq:pair_sauter_massive}.

\begin{figure}[h]
\begin{center}
 \begin{minipage}{0.46\textwidth}
 \includegraphics[width=1.0\textwidth]{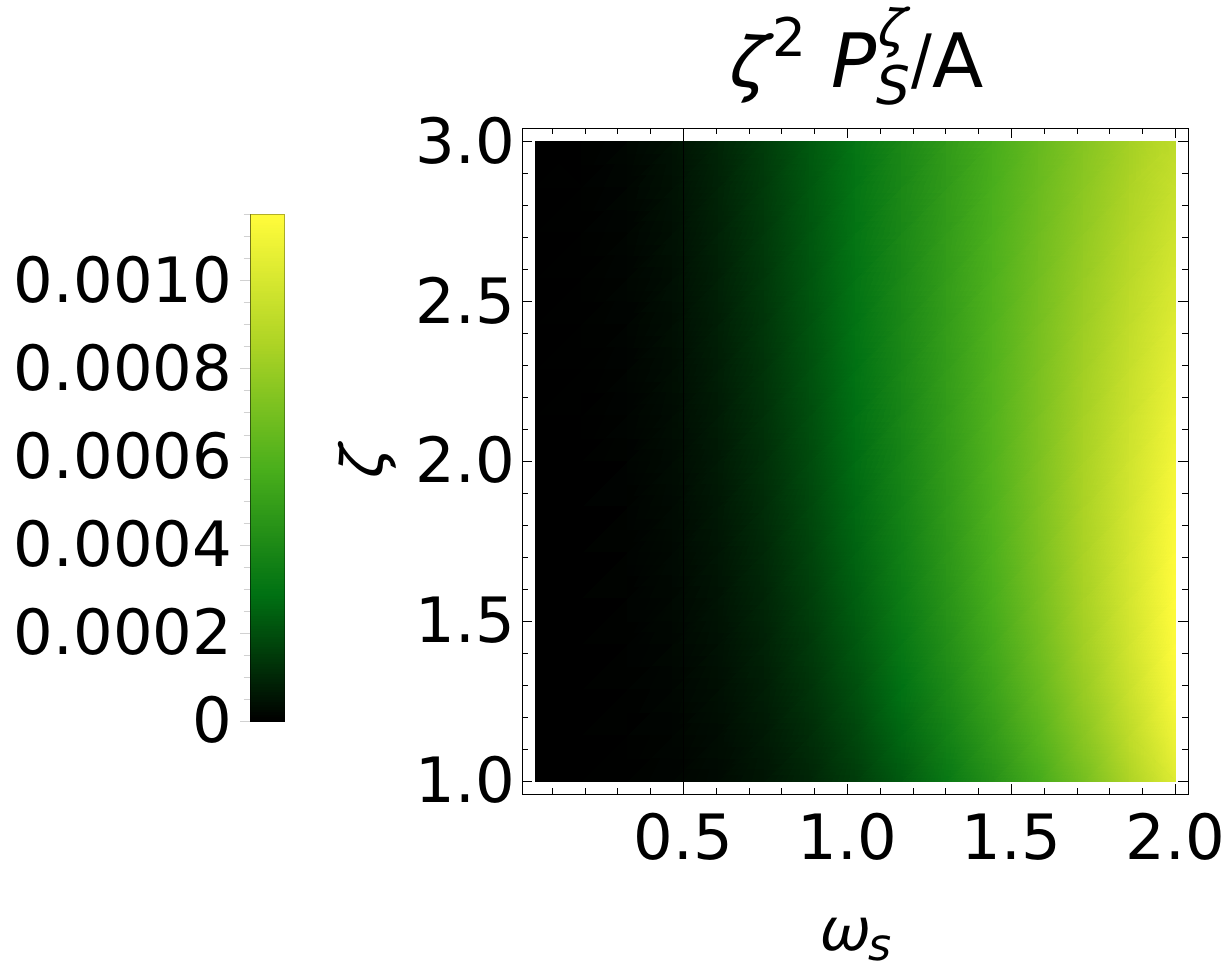} 
 \end{minipage}
\hspace{0.02\textwidth}
\begin{minipage}{0.46\textwidth}
 \vspace{0.5cm}\includegraphics[width=1.0\textwidth]{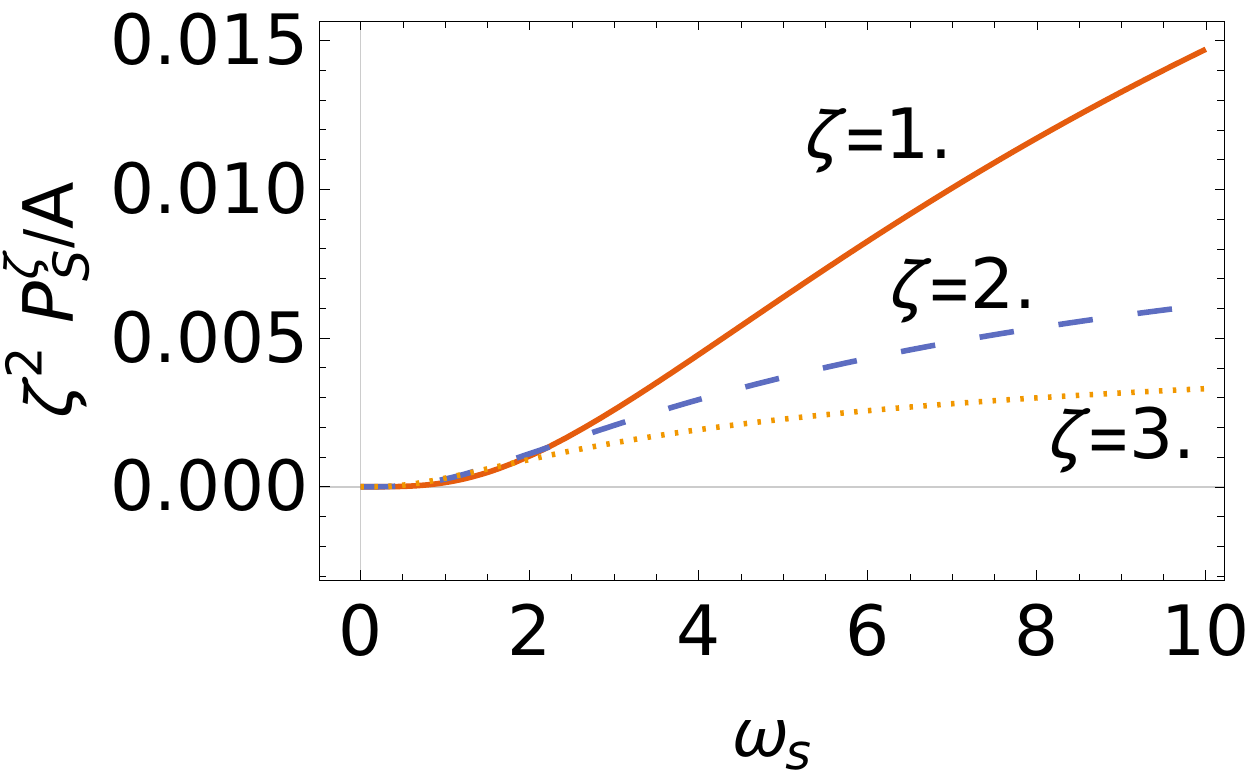} 
 \end{minipage}
 \end{center}
 \label{fig:sauter}
 \caption{Probability of assisted particle production for a Sauter pulse per unit area, $\ampli^2 P_{S}^\ampli/A$.
 The left panel shows its density plot as a function of the frequency $\omega_S$ and the background $\ampli$. 
 The right panel displays its behaviour as the frequency varies, considering different backgrounds. 
 We have set the amplitude of the pulse ($\eta_0$) to unity.}
 \end{figure}
 
 We can also consider the massless case together with an arbitrary  $\ampli$. 
An expansion of the pair production probability for large and small frequencies thus gives
 \begin{align}\label{eq:pair_sauter_assisted}
 \frac{ P_S^{\ampli}}{A}=
 \eta_0^2 \omega_S^4
 \times \begin{cases}
 \frac{7 }{ \pi ^{9}} \Big[  \zeta_{\rm R} (7)- \frac{54}{7\pi^2} \zeta_{\rm R} (9) (\omega_S\ampli)^2 +\cdots\Big],\quad \ampli\omega_S\ll 1 ,
  \\
 \frac{1}{(\ampli\omega_S)^4}
 \Big[ 
 \frac{12 \zeta_{\rm R} (3)}{\pi ^5 } - \frac{8}{3 \pi  \zeta\omega_S}+\cdots
 \Big]
 ,\quad \ampli\omega_S\gg 1.
 \end{cases}
\end{align}
In particular, for vanishing $\ampli$ we recover the expected massless limit of Eq.~\eqref{eq:pair_sauter_massive}. 
Increasing the value of $\ampli$ gives rise to an anti-assisted effect, 
contrary to the one found for electromagnetic backgrounds for Schwinger pair production~\cite{Schutzhold:2008pz, Linder:2015vta}.
Naively, since a constant potential alone does not contribute to pair creation in our scalar setup\footnote{This is one intrinsic difference with the electromagnetic case, where for example constant electric fields do produce pairs, giving  rise to the Sauter--Schwinger effect.}, one could have expected that its addition would have had no effect on pair creation. 
Instead, the nonlocal feature of the generated form factor is nontrivial.
The fact that it diminishes the pair production can be physically understood from the fact that a larger $\ampli$ tends to repel the quantum field, 
as explained in Sec.~\ref{sec:ea_inho}.

Lastly, taking into account the discourse developed in Sec.~\ref{sec:ea_pure}, 
we can analyze the case in which we rescale the amplitude of the perturbation, 
$\eta_0=:\ampli\tilde\eta $, such that $\tilde \eta$ is taken to be small. 
At the practical level, this is analogous to saying that we multiply the pair production probability by $\ampli^2$.
We have plotted in the left panel of Fig.~\ref{fig:sauter} a density plot of the rescaled probability as a function of both $\omega_S$ and $\ampli$;
one can see that for $\omega_S\lesssim 1.5$, increasing the coupling $\ampli$ leads to larger probabilities. 
The right panel of Fig.~\ref{fig:sauter} provides a clearer picture of this effect, 
showing the rescaled probability as a function of $\omega_S$ for three distinct values of $\ampli\sim \mathcal{O}(1)$.
For $\omega_S\sim 1.5$ we see that the hierarchy between the curves is inverted.

\section{Conclusions}\label{sec:conclusion}

In the present article we have studied the problem of a quantum scalar field theory with local semitransparent Neumann boundary conditions on a plate, which can also be understood as interactions with an external (classical) field confined to the latter.

First of all, we have shown that the quantum contributions to the effective action can be understood in terms of quantities in the phase space of a first quantization.
Worldline techniques are by far much more developed in the case of configuration space~\cite{Schubert:2001he}, rendering the study of such path integrals \emph{per se} an interesting problem.

Second, taking into account the well-known fact that the studied interaction requires regularization, we have devised a regularization appropriate to our path-integral methods. We have shown that in this way we are able to rederive previously obtained results for the homogeneous case. 
Notice that this regularization provides also a physical interpretation in terms of an effective finite width of the plate, since we have to evaluate the intermediate heat-kernels at noncoincident points.

These technical developments allowed the computation of the heat-kernel at quadratic level on the perturbations $\eta$ around a homogeneous background field $\ampli$ and the corresponding effective action in the $\dime\equiv 4$ case. Notably, these results are exact in $\ampli$; 
they are (generally) given by a nonlocal operator acting on the perturbations $\eta$,
what can be seen from the appearance of nonanalytic contributions.

Such nonanalytic contributions are responsible for the pair production that we have encountered in Minkowski spacetime, where the perturbations are used to model dynamical properties of the plates, very much akin to the situation in the Dynamical Casimir effect. In this scenario we have considered two possibilities, 
a harmonic excitation and a Sauter pulse. 
For large masses, the Sauter background displays an exponential cutoff, which is softer than the Heaviside present for the harmonic pulse.
In the massless Sauter case, 
the introduction of a background $\ampli$ leads to a scenario of anti-assisted pair production.
This effect can be inverted in a region of the parameters space if the amplitude of the pulse scales linearly with $\ampli$, what is possible taking into account the discussion in Sec.~\ref{sec:ea_pure} regarding the smallness of $\eta$ and $\ampli$.

A special comment deserves the finding of a duality between the field theory obeying local semitransparent Neumann boundary conditions and the equivalent Dirichlet one.
Indeed, we have shown that there exists a mapping between the relevant heat-kernels, 
which becomes an exact strong/weak duality at the level of the corresponding renormalized quantum effective actions.
Notice that this is a new duality, different from the Fermi--Bose duality discussed in condensed matter by means of the Girardeau mapping~\cite{Girardeau:1960,Cheon:1998iy,Granet:2021buv}. 
To clarify this point, first notice that the latter connects the Lieb--Liniger model~\cite{Lieb:1963rt} and the Cheon--Shigehara one, which are both one-dimensional models, while our results are valid in $D$-dimensions. 
Second, as discussed in Sec.~\ref{sec:hk_duality}, we impose no type of symmetry under the parity transformation $\tilde x\to -\tilde x$ and we always work with a scalar bosonic field, contrary to the change in statistics of the Girardeau mapping.
Additionally, we don't introduce a self-interaction; instead, our field interacts with a spacetime-dependent background potential $\eta$.
Nevertheless, taking into account the several experimental accomplishments based on the Fermi--Bose duality~\cite{Wilson:2020,Paredes:2004}, it will be of interest to explore possible experimental roads of the newly devised duality.

Regarding possible future developments, it will be interesting to try to generalize our results to the problem where nonlocal boundary conditions are imposed. This may provide a way to analyze the appearance of topological effects.

One further peculiarity of the interaction considered in this article is that it can be thought as the first term in an effective field theory expansion, which has undergone a thin-shell limit.
As such, it may find applications in trying to understand the nature of dark matter. Indeed,
one open possibility is that it may behave as a field with  unusual couplings~\cite{Fichet:2017bng,Brax:2022wrt}.

Finally, another useful advance would be the development of a numerical code to tackle the problem discussed here. The situation is more involved than the usual cases, since a naive  adaptation of  the numerical Worldline techniques~\cite{Franchino-Vinas:2019udt, Gies:2006cq, Gies:2006bt} to phase space suffer from the so-called sign problem.

%%%%%%%%%%%%%%
%%%%%%%%%%%%%%
%%%%%%%%%%%%%%
%%%%%%%%%%%%%%
%%%%%%%%%%%%%%

%%%%%%%%%%%%%%
%%%%%%%%%%%%%%
%%%%%%%%%%%%%%
%%%%%%%%%%%%%%
%%%%%%%%%%%%%%

\section*{Acknowledgements}
S.A.F. and L.M. acknowledge support from Project 11/X748, UNLP. S.A.F. was supported by Helmholtz-Zentrum Dresden-Rossendorf (HZDR) and Consejo
Nacional de Investigaciones Cient\'ificas y T\'ecnicas (CONICET). F.D.M was supported by Agencia Nacional de
Promoción Cient\'ifica y Tecnol\'ogica (ANPCyT), Consejo
Nacional de Investigaciones Cient\'ificas y T\'ecnicas (CONICET), and Universidad Nacional de Cuyo (UNCuyo).

%%%%%%%%%%%%%%
%%%%%%%%%%%%%%

\appendix

\section{The partition function in the Worldline}\label{app:vev}
In this appendix we will recall how to compute the partition function for an open scalar line in the Worldline Formalism, 
considering a source $k$ for $\dot{x}$:
\begin{align}
 \begin{split}
Z_{\dot x}[k]:&=\frac{\pathi[\,e^{-\int_{t_1}^{t_2}\dx[t] \frac{\dot{x}^2(t)}{4} +\int_{t_1}^{t_2} \dx[t]\, k(t) \dot{x}(t) }][x][t_1][x'][t_2]}
{\pathi[\,e^{-\int_{t_1}^{t_2}\dx[t] \frac{\dot{x}^2(t)}{4} }][x][t_1][x'][t_2]} 
  .
 \end{split}
\end{align}
First of all, we may perform a change of variable in the path, so that we get a path integral over trajectories that obey initial and final Dirichlet boundary conditions.
In order to do so, we split the path into a  classical and a quantum contribution; substituting $x(t)\to y(t)+(x'-x)(t-t_1)/(t_2-t_1)+x$ we get
\begin{align}
 \begin{split}
Z_{\dot x}[k]
 &=\frac{\pathi[\,e^{-\int_{t_1}^{t_2}\dx[t] \frac{\left(\dot{y}(t)+\frac{x'-x}{t_2-t_1}\right)^2}{4} +\int_{t_1}^{t_2} \dx[t]\, k(t) \left(\dot{y}(t)+\frac{x'-x}{t_2-t_1}\right) }][0][t_1][0][t_2][y]
 }
{\pathi[\,e^{-\int_{t_1}^{t_2}\dx[t] \frac{\dot{x}^2(t)}{4} }][x][t_1][x'][t_2]} 
\\
 &= e^{\frac{x'-x}{t_2-t_1} \int_{t_1}^{t_2} \dx[t]\, k(t) } \frac{\pathi[\,e^{-\int_{t_1}^{t_2}\dx[t] \frac{\dot{y}^2(t)}{4} -\int_{t_1}^{t_2} \dx[t]\, \dot{k}(t) {y}(t) }][0][t_1][0][t_2][y]
  }
{\pathi[\,e^{-\int_{t_1}^{t_2}\dx[t] \frac{\dot{y}^2(t)}{4} }][0][t_1][0][t_2][y]} ,
 \end{split}
\end{align}
where in the last line we have performed an integration by parts to get rid of the derivatives acting on $y(t)$ in the source term.

We have thus reduced our problem to the computation of the partition function with Dirichlet boundary conditions, 
which can be readily solved by inverting the kinetic term.
Indeed, a straightforward computation gives
\begin{align}
 \frac{\pathi[\,e^{-\int_{t_1}^{t_2}\dx[t] \frac{\dot{y}^2(t)}{4} -\int_{t_1}^{t_2} \dx[t]\, \dot{k}(t) {y}(t) }][0][t_1][0][t_2][y]
  }
{\pathi[\,e^{-\int_{t_1}^{t_2}\dx[t] \frac{\dot{y}^2(t)}{4} }][0][t_1][0][t_2][y]} 
 = e^{\frac{1}{4}\int_{t_1}^{t_2}\dx[t] \int_{t_1}^{t_2}\dx[s] \,\dot{k}(s) G(s,t) \dot{k}(t) } ,
\end{align}
where the required symmetric Green function  is defined as
\begin{align}
 G(s,t):= \frac{4}{(t_2-t_1)} (s-t_1)(t_2-t),\quad t_1<s<t<t_2.
\end{align}
This Green function satisfies as customarily the differential equation
\begin{align}
 -\frac{1}{4}\partial^2_s G(s,t)=\delta(s-t),
\end{align}
as well as the boundary conditions $G(0,t)=G(t_2,t)=0=G(s,0)=G(s,t_2)$.

One can further simplify the expression in the present case integrating by parts in the exponent; 
explicitly employing the boundary conditions satisfied by the Green function, we get
\begin{align}
 \frac{\pathi[\,e^{-\int_{t_1}^{t_2}\dx[t] \frac{\dot{y}^2(t)}{4} -\int_{t_1}^{t_2} \dx[t]\, \dot{k}(t) {y}(t) }][0][t_1][0][t_2][y]
  }
{\pathi[\,e^{-\int_{t_1}^{t_2}\dx[t] \frac{\dot{y}^2(t)}{4} }][0][t_1][0][t_2][y]} 
 = e^{\frac{1}{4}\int_{t_1}^{t_2}\dx[t] \int_{t_1}^{t_2}\dx[s] \,{k}(s) \partial_s\partial_t G (s,t) {k}(t) },
\end{align}
%where the dots on the left (right) of the Green function should be intented respectively as partial derivatives on the left (right) variable.
A direct computation shows that the second partial derivative involved in the computation is given by
\begin{align}
 \partial_s\partial_t G(s,t) =-\frac{4}{(t_2-t_1)}\left[1-\delta(t-s)(t_2-t_1) \right].
\end{align}
Adding all these results toghether we are led to our final expression
\begin{align}\label{eq:partition_function}
 \begin{split}
Z_{\dot x}[k]
 &= e^{\frac{x'-x}{t_2-t_1} \int_{t_1}^{t_2} \dx[t]\, k(t)  -\frac{1}{(t_2-t_1)} \left( \int_{t_1}^{t_2}\dx[t] \,{k}(t)\right)^2+\int_{t_1}^{t_2} \dx[t]\, k^2(t) } .
 \end{split}
\end{align}

\section{Intermediate-time integrals of chained heat-kernels}\label{app:integrals}

In Sec. \ref{sec:worldline_hk}, we are lead to expressions that involve integrals of chains of free heat-kernels in the intermediate time $t$, 
combined with increasing negative powers of that time and its distance to the end time $T$. 
One can compute the first of them by considering the first order in $\ampli$ of the result~\cite[Eq. $(16)$]{Franchino-Vinas:2020okl}:
\begin{align}\label{eq:integral_HK_chain}
 \int_0^T \dx[t] 
 \frac{e^{-\frac{a}{t}-\frac{b}{T-t}}}{\sqrt{t} \sqrt{T-t}}&= \pi  \text{erfc}\left(\frac{\sqrt{a}+\sqrt{b}}{\sqrt{T}}\right).
\end{align}
The algorithm to obtain integrals with higher powers in the intermediate time involves differentiating this expression in terms of $a$ or $b$. 
As a matter of completion, we mention the first of them:
\begin{align}
 \int_0^T \dx[t] \frac{e^{-\frac{a}{t}-\frac{b}{T-t}}}{\sqrt{t} (T-t)^{3/2}}&= \frac{\sqrt{\pi } e^{-\frac{\left(\sqrt{a}+\sqrt{b}\right)^2}{T}}}{\sqrt{b} \sqrt{T}},
 \\
 \int_0^T \dx[t] \frac{e^{-\frac{a}{t}-\frac{b}{T-t}}}{\sqrt{t} (T-t)^{5/2}}&= \frac{\sqrt{\pi } e^{-\frac{\left(\sqrt{a}+\sqrt{b}\right)^2}{T}} \left(2 \sqrt{a} \sqrt{b}+2 b+T\right)}{2 b^{3/2} T^{3/2}},
 \\
 \int_0^T \dx[t] \frac{e^{-\frac{a}{t}-\frac{b}{T-t}}}{t^{3/2} \sqrt{T-t}}&= \frac{\sqrt{\pi } e^{-\frac{\left(\sqrt{a}+\sqrt{b}\right)^2}{T}}}{\sqrt{a} \sqrt{T}},
 \\
 \int_0^T \dx[t] \frac{e^{-\frac{a}{t}-\frac{b}{T-t}}}{t^{3/2} (T-t)^{3/2}}&= \frac{\sqrt{\pi } \left(\sqrt{a}+\sqrt{b}\right) e^{-\frac{\left(\sqrt{a}+\sqrt{b}\right)^2}{T}}}{\sqrt{a} \sqrt{b} T^{3/2}},
 \\
 \int_0^T \dx[t] \frac{e^{-\frac{a}{t}-\frac{b}{T-t}}}{t^{3/2} (T-t)^{5/2}}&= \frac{\sqrt{\pi } e^{-\frac{\left(\sqrt{a}+\sqrt{b}\right)^2}{T}} \left(\sqrt{a} (4 b+T)+2 a \sqrt{b}+2 b^{3/2}\right)}{2 \sqrt{a} b^{3/2} T^{5/2}},
 \\
 \int_0^T \dx[t]  \frac{e^{-\frac{a}{t}-\frac{b}{T-t}}}{t^{5/2} \sqrt{T-t}}&= \frac{\sqrt{\pi } e^{-\frac{\left(\sqrt{a}+\sqrt{b}\right)^2}{T}} \left(2 \sqrt{a} \sqrt{b}+2 a+T\right)}{2 a^{3/2} T^{3/2}},
 \\
 \int_0^T \dx[t] \frac{e^{-\frac{a}{t}-\frac{b}{T-t}}}{t^{5/2} (T-t)^{3/2}}&=                                                                                                                                                                                               \frac{\sqrt{\pi } e^{-\frac{\left(\sqrt{a}+\sqrt{b}\right)^2}{T}} \left(2 \sqrt{a} \left(\sqrt{a}+\sqrt{b}\right)^2+\sqrt{b} T\right)}{2 a^{3/2} \sqrt{b} T^{5/2}},
 \\
 \int_0^T \dx[t] \frac{e^{-\frac{a}{t}-\frac{b}{T-t}}}{t^{5/2} (T-t)^{5/2}}&= \frac{\sqrt{\pi } e^{-\frac{\left(\sqrt{a}+\sqrt{b}\right)^2}{T}} \left(T \left(a^{3/2}+b^{3/2}\right)+2 \sqrt{a} \sqrt{b} \left(\sqrt{a}+\sqrt{b}\right)^3\right)}{2 a^{3/2} b^{3/2} T^{7/2}}.
\end{align}

\section{Integrals involving Hermites , gaussians and inverse powers of the intermediate time}\label{app:integral_hermite}
A crucial step in obtaining an all-order expression for the heat-kernel in Sec.~\ref{sec:worldline_hk}, is the derivation of a closed expression for the integral
\begin{align}
 \begin{split}\label{eq:integral_hermites0}
A_{n,m}(x,y):=&\int_0^{1} \dx[t] \frac{e^{-\frac{x^2}{4t}}}{t^{(1+n)/2}}\frac{e^{-\frac{y^2}{4(1-t)}}}{(1-t)^{(1+m)/2}} H_{n}\left(\frac{|x|}{2\sqrt{t}}\right) H_m\left(\frac{|y|}{2\sqrt{1-t}}\right). \end{split}
\end{align}
Employing the wellknown formula to generate the Hermite polynomials as derivatives of a Gaussian,
\begin{align}\label{eq:hermite_as_derivative}
 H_n(x)=(-1)^n e^{x^2}\partial^n_x e^{-x^2},
\end{align}
we can recast Eq. \eqref{eq:integral_hermites0} as
\begin{align}
 \begin{split}\label{eq:integral_hermites}
A_{n,m}(x,y)
  &=(-2)^{n+m} \operatorname{sign}^n(x)\operatorname{sign}^m(y) \partial_x^n\partial_y^m \int_0^{1} \dx[t]  \frac{1}{t^{1/2}}  \frac{1}{(1-t)^{1/2}}  e^{-\frac{x^2}{4t}}  e^{-\frac{y^2}{4(1-t)}} 
 \\
 &=(-2)^{n+m} \operatorname{sign}^n(x)\operatorname{sign}^m(y) \partial_x^n\partial_y^m 
 \pi  \text{erfc}\left(\frac{\sqrt{x^2}+\sqrt{y^2}}{2}\right),
 \end{split}
\end{align}
where in the last line we have employed our result~\eqref{eq:integral_HK_chain}.
If $x>0$, then we may set $\sqrt{x^2}=x$ in Eq. \eqref{eq:integral_hermites}, since the derivatives act by definition only locally. 
If also $y>0$, employing once more the generating formula \eqref{eq:hermite_as_derivative} we find 
\begin{align}
 \begin{split}\label{eq:integral_hermites2}
A_{n,m}(x,y)
 &=2 \sqrt{\pi} e^{-\frac{(x+y)^2}{4}} H_{n+m-1}\left(\frac{x+y}{2}  \right).
 \end{split}
\end{align}
Extending this analysis to any sign of $x$ and $y$, and allowing also for a final time $T$ different from one, we prove the following relation
\begin{align}
 \begin{split}\label{eq:integral_hermites_final}
&\int_0^{T} \dx[t] \frac{e^{-\frac{x^2}{4t}}}{t^{(1+n)/2}}\frac{e^{-\frac{y^2}{4(T-t)}}}{(1-t)^{(1+m)/2}} H_{n}\left(\frac{|x|}{2\sqrt{t}}\right) H_m\left(\frac{|y|}{2\sqrt{T-t}}\right)
 \\
 &\hu\hu\hu\hu=2 \sqrt{\pi} \frac{e^{-\frac{(x+y)^2}{4T}}}{T^{(m+n)/2}} H_{n+m-1}\left(\frac{|x|+|y|}{2\sqrt{T}}  \right),
 \end{split}
\end{align}
valid for\footnote{Notice that if $x=0$ or $y=0$, then the departing integral in Eq. \eqref{eq:integral_hermites} will be in general ill-defined. 
This is not an obstacle to the computations in the body of this article, given that we always work in a regularized framework, 
in which 
we have to consider the limiting cases $x,\,y\to0$. }
$x,\,y\in\mathbb{R}-\{0\}$ and $n,\,m=0,\,1,\cdots$,  if Hermite's polynomials with negative index are understood in terms of parabolic cylinder functions.

\section{Series of Hermite polynomials}\label{app:series_hermite}
In Sec.~\ref{sec:worldline_hk}, the resummation of the heat-kernel's $\ampli$-expansion involves the computation of a series of Hermite functions, 
\begin{align}
 T(x,\beta):&=\sum_{n=0}^{\infty} H_n(x) \beta^n.
\end{align}

A way to obtain a closed expression for this series is to notice that, by using  \eqref{eq:hermite_as_derivative}, we have
\begin{align}
 \begin{split}
T(x,\beta)&=\sum_{n=0}^{\infty}  e^{x^2} (-\beta)^n \frac{{\rm d}^n}{{\rm d} x^n}e^{-x^2}
 \\
 &=e^{x^2} \frac{1}{1+ \beta \frac{{\rm d}}{{\rm d} x}} e^{-x^2},
 \end{split}
\end{align}
where we have employed the formal expression of the geometric series. 
Solving it for $T$ we arrive at the following differential equation
\begin{align}
 \begin{split}
\left({1+ \beta \frac{{\rm d}}{{\rm d} x}}\right) e^{-x^2} T(x,\beta)=  e^{-x^2},
 \end{split}
\end{align}
whose more general solution is given by 
\begin{align}
 \tilde T(x,\beta)&=c_1(\beta) e^{x^2-\frac{x}{\beta }}+\frac{\sqrt{\pi }}{2 \beta } e^{\frac{1}{4 \beta ^2}+x^2-\frac{x}{\beta }} \text{erf}\left(x-\frac{1}{2 \beta }\right).
\end{align}
The ``constant of integration'' $c_1$ may be fixed by analyzing the behaviour of $T$ for small $\beta$, from which we get 
\begin{align}
 c_1(\beta)=-\frac{\sqrt{\pi } e^{\frac{1}{4 \beta ^2}}}{2 \beta }.
\end{align}
This implies that the desired function is given by 
\begin{align}
 T(x,\beta)=-\frac{\sqrt{\pi } e^{\frac{(1-2 \beta  x)^2}{4 \beta ^2}} \text{erfc}\left(x-\frac{1}{2 \beta }\right)}{2 \beta }.
\end{align}

\printbibliography

\end{document}